\documentclass[5p]{elsarticle}
\biboptions{round, authoryear}
\usepackage[utf8]{inputenc}
\usepackage{amsmath}
\usepackage{epstopdf}
\usepackage{multirow}
\usepackage{subfig}
\usepackage{threeparttable}
\usepackage{url}

\begin{document}
\begin{frontmatter}
\title{Mapping of Jupiter's tropospheric NH$_3$ abundance using ground-based IRTF/TEXES observations at 5 $\mu$m}

\author[obspm]{Doriann Blain\corref{cor1}}
\ead{doriann.blain@obspm.fr}
\author[obspm]{Thierry Fouchet}
\author[swri]{Thomas Greathouse}
\author[obspm]{Thérèse Encrenaz}
\author[obspm]{Benjamin Charnay}
\author[obspm]{Bruno Bézard}
\author[jpl]{Cheng Li}
\author[obspm]{Emmanuel Lellouch}
\author[jpl]{Glenn Orton}
\author[ul]{Leigh N. Fletcher}
\author[obspm]{Pierre Drossart}

\address[obspm]{LESIA, Observatoire de Paris, PSL Research University, CNRS, Sorbonne Université, Univ. Paris Diderot, Sorbonne Paris Cité, 92195, Meudon, France}
\address[swri]{SwRI, Div. 15, San Antonio, TX 78228, USA}
\address[jpl]{Jet Propulsion Laboratory, California Institute of Technology, Pasadena, CA 91109, USA}
\address[ul]{Department of Physics and Astronomy, University of Leicester, University Road, Leicester, LE1 7RH, UK}

\cortext[cor1]{Corresponding author.}

\begin{abstract}
We report on results of an observing campaign to support the Juno mission. At the beginning of 2016, using TEXES (Texas Echelon cross-dispersed Echelle Spectrograph), mounted on the NASA Infrared Telescope Facility (IRTF), we obtained data cubes of Jupiter in the 1930--1943 cm$^{-1}$ spectral ranges (around 5 $\mu$m), which probe the atmosphere in the 1--4 bar region, with a spectral resolution of $\approx$ 0.15 cm$^{-1}$ and an angular resolution of $\approx$ 1.4''.
This dataset is analysed by a code that combines a line-by-line radiative transfer model with a non-linear optimal estimation inversion method. The inversion retrieves the vertical abundance profiles of NH$_3$ --- which is the main contributor at these wavelengths --- with a maximum sensitivity at $\approx$ 1--3 bar, as well as the cloud transmittance. This retrieval is performed on more than one thousand pixels of our data cubes, producing maps of the disk, where all the major belts are visible.
We present our retrieved NH$_3$ abundance maps which can be compared with the distribution observed by Juno's MWR \citep{Bolton2017, Li2017} in the 2 bar region and discuss their significance for the understanding of Jupiter's atmospheric dynamics.
We are able to show important latitudinal variations --- such as in the North Equatorial Belt (NEB), where the NH$_3$ abundance is observed to drop down to 60 ppmv at 2 bar --- as well as longitudinal variability. In the zones, we find the NH$_3$ abundance to increase with depth, from 100 $\pm$ 15 ppmv at 1 bar to 500 $\pm$ 30 ppmv at 3 bar. We also display the cloud transmittance--NH$_3$ abundance relationship, and find different behaviour for the NEB, the other belts and the zones. Using a simple cloud model \citep{Lacis1974, Ackerman2001}, we are able to fit this relationship, at least in the NEB, including either NH$_3$-ice or NH$_4$SH particles with sizes between 10 and 100 $\mu$m.
\end{abstract}
\end{frontmatter}

\section{Introduction}
Ammonia (NH$_3$) is an important molecule for the understanding of Jupiter's atmosphere. Indeed, the most commonly accepted models of Jupiter's atmosphere, such as from \cite{Atreya1999} show that the condensation of NH$_3$ plays a major role in the presumed cloud structure of Jupiter. It is assumed to react with hydrogen sulfide (H$_2$S) to form an ammonium hydrosulfide (NH$_4$SH) cloud at around 2 bar, and to condense to form a NH$_3$-ice cloud at about 0.8 bar. 

\begin{figure}[ht!]
\centering
\includegraphics[width=0.3\textwidth]{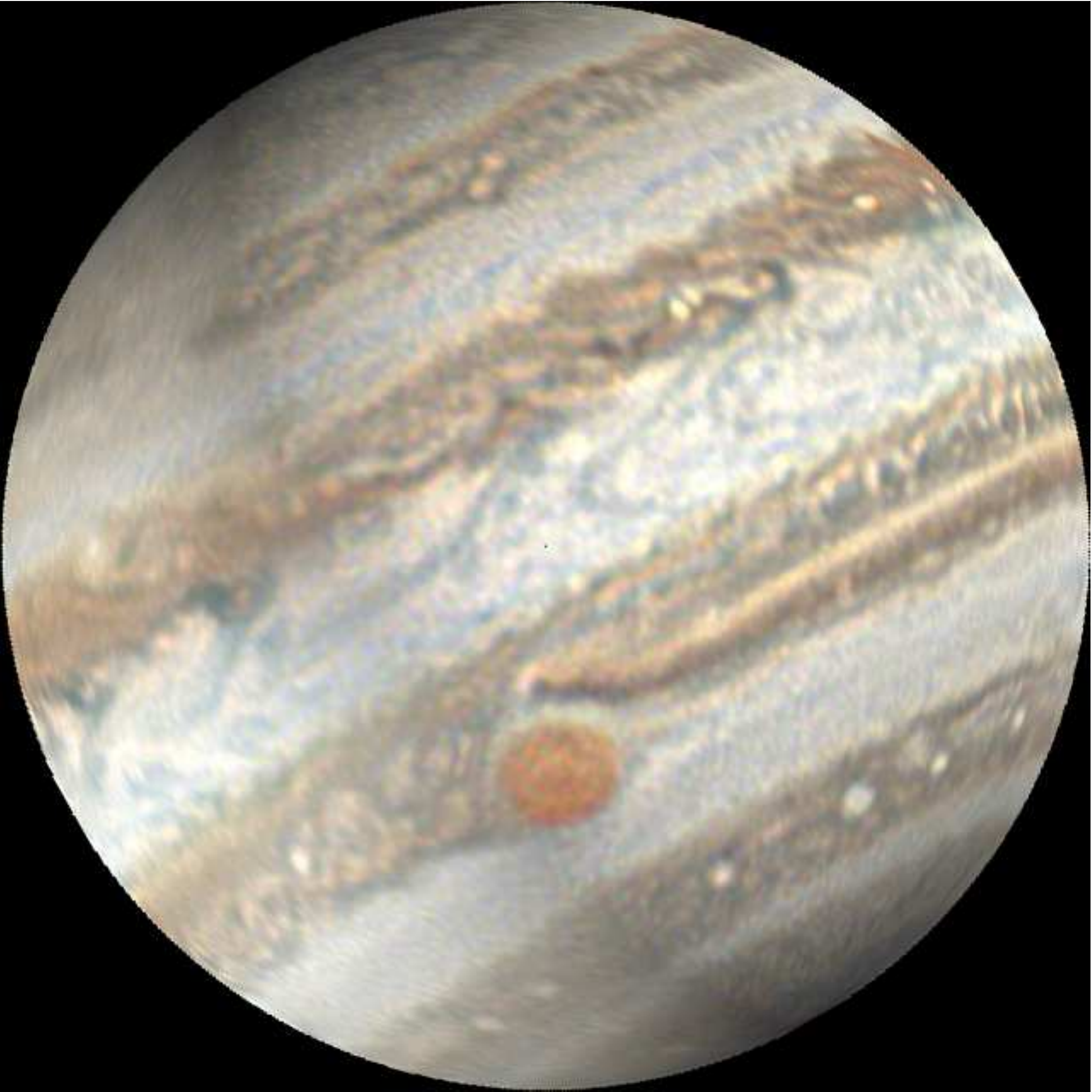}
\caption{\label{fig:disk_visible_light}Reconstitution of the apparent disk of Jupiter the 16 January 2016, 11:10 UT. Orthographic projection of a mosaic of pictures taken by \citet{Einaga2016} between 15 and 16 January 2016 UT with a 300 mm Newton telescope from Kasai-City, Hyogo-Prefecture, Japan.}
\end{figure}

Observations made at radio wavelengths by \citet{dePater1986, dePater2016} for example, are consistent with this model. The NH$_3$ abundance profile is found to sharply decrease above the 0.6-bar and 2-bar pressure levels --- with variations between the North Equatorial Belt (NEB) and the Equatorial Zone (EZ) ---, where the hypothetical clouds are predicted to lie, while the abundance remains constant below the 2-bar pressure level. In contrast to these observations and theory, the Galileo atmospheric probe measurements of the NH$_3$ volume mixing ratio (VMR) showed NH$_3$ to increase down to 8 bar \citep{Folkner1998, Sromovsky1998}. However, the probe entered a specific region called a hotspot --- a small, bright region at 5 $\mu$m ---, and several remote sensing measurements \citep{Fouchet2000, dePater2001, Bezard2002, Bjoraker2015, Fletcher2016} have demonstrated that hotspots are not representative of the whole atmosphere.

More recently, the first published results of the Juno Microwave Radiometer (MWR) \citep{Bolton2017, Li2017} undeniably show that the distribution of NH$_3$ below the 1-bar pressure level is much more complex than previously thought, with strong variations observed both with altitude and with latitude --- though this complexity was previously observed above the 1-bar pressure level from 10-$\mu$m observations \citep{Achterberg2006, Fletcher2016}. Most strikingly, MWR profiles display a minimum with altitude at about 7 bar everywhere on the planet except in the EZ. In addition, \cite{Orton2017} have highlighted a potential inconsistency at some latitudes between JIRAM radiance at 5 $\mu$m and MWR brightness temperature that has yet to be understood. Moreover, MWR currently published results may not be representative of the whole planet. Indeed, MWR channels have a full width half maximum (FWHM) footprint from 2$^\circ$ at the equator to 20$^\circ$ near the poles, so even if MWR data cover latitudes pole to pole, only a narrow longitude range is explored during each perijove pass.

Ammonia could also play a role in the visible-light appearance of Jupiter, itself correlated with the vertical and horizontal winds in the troposphere. The NH$_3$-ice clouds could be responsible for the white color of Jupiter's zones in the visible, and are associated with intense vertical updrafts bringing NH$_3$ to high altitudes \citep{Owen1981}. However, this relation between NH$_3$ and the visible brightness of Jupiter has been tempered by the work of \cite{Giles2015}, which showed that pure NH$_3$-ice clouds are not consistent with Cassini VIMS 5-$\mu$m data. These results are in agreement with \citet{Baines2002} observations using the Galileo Near-Infrared Mapping Spectrometer (NIMS), which showed that pure NH$_3$-ice clouds have bee identified on less than 1$\%$ of the area observed in the study. In the same work, to explain this discrepancy between the observations and thermodynamic predictions, it was suggested that NH$_3$-ice material might be altered either by photochemistry or coated by another material. The brown color of the belts may be due to a deeper cloud deck possibly containing a sulfide, like NH$_4$SH, as evoked by \citet{Owen1981}. The color of the Great Red Spot (GRS), might be due to NH$_3$ photodissociation byproducts reacting with acetylene (C$_2$H$_2$) at high altitude \citep{Carlson2016}, or irradiation of NH$_4$SH particle \citep{Loeffler2018}. In all cases, NH$_3$ seems to play a role in the cloud formation --- and therefore, the colors in visible-light --- of Jupiter. Joint cloud opacity and NH$_3$ gas abundance measurement could therefore be useful to assess the importance of this role.

In this work, we use high-resolution spectral cubes, described in section~\ref{sec:observations}, covering most of Jupiter's disk to simultaneously retrieve the abundance of NH$_3$ in the troposphere and the cloud transmittance. In section~\ref{sec:methods} we present our methodology, and our results are described and discussed in section~\ref{sec:results}. We display a map of the NH$_3$ abundance in the troposphere, covering a wide range of longitudes and latitudes. We aim to both offer a complementary view of MWR data and try to spot localised features that may be missed by MWR due to its narrow longitudinal coverage.

\begin{figure}[t]
\centering
\includegraphics[width=0.48\textwidth]{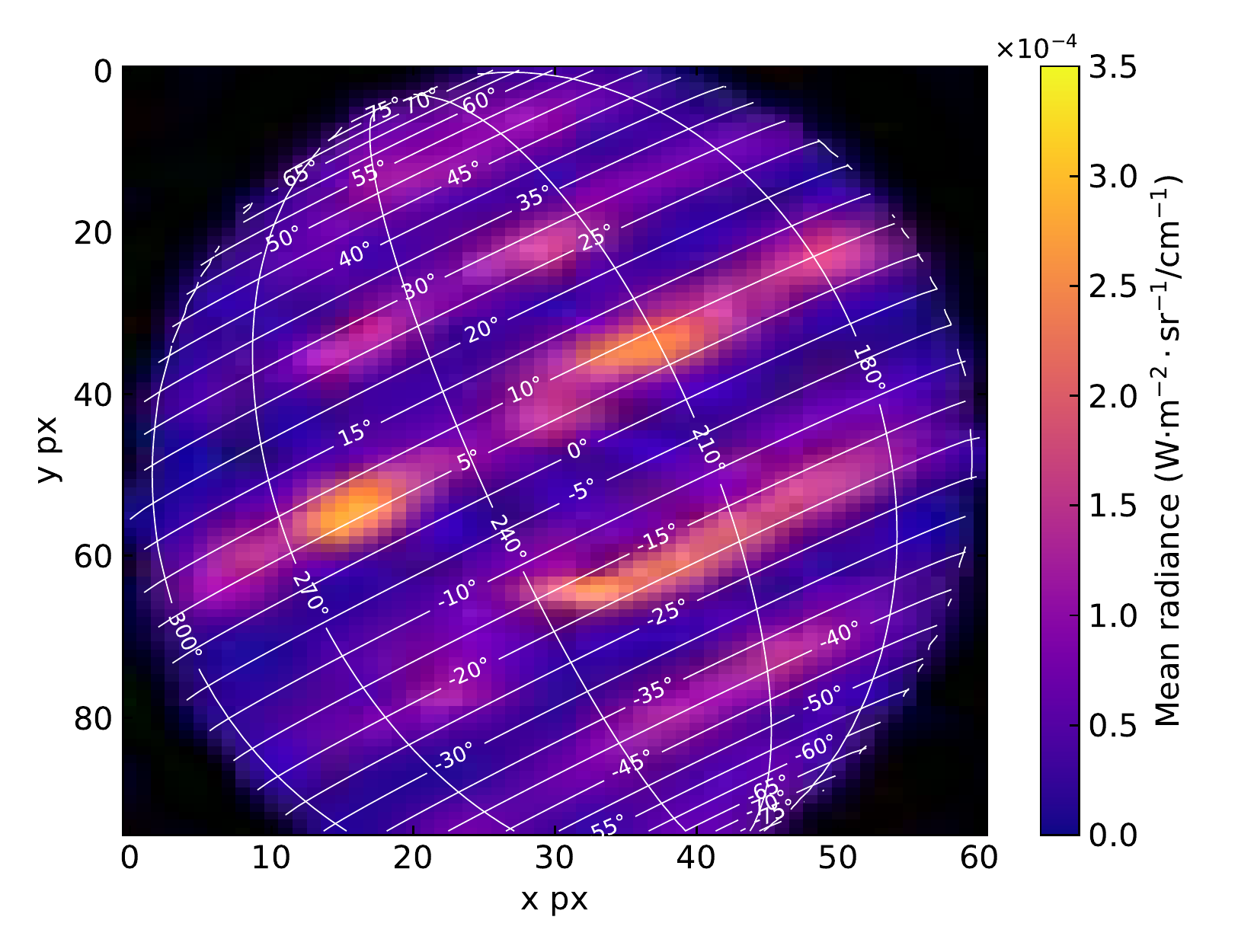}
\caption{\label{fig:disk_texes}Map of the mean observed radiance for the reduced spectral cube in the 1930--1943 cm$^{-1}$ wavenumber range taken the 16 January 2016 at 11h10 UTC. The Great Red Spot can be seen around pixel [32, 70]. The longitudes are in system III and the latitudes are planetocentric.}
\end{figure}

\begin{figure*}[p]
\includegraphics[width=0.9\textwidth]{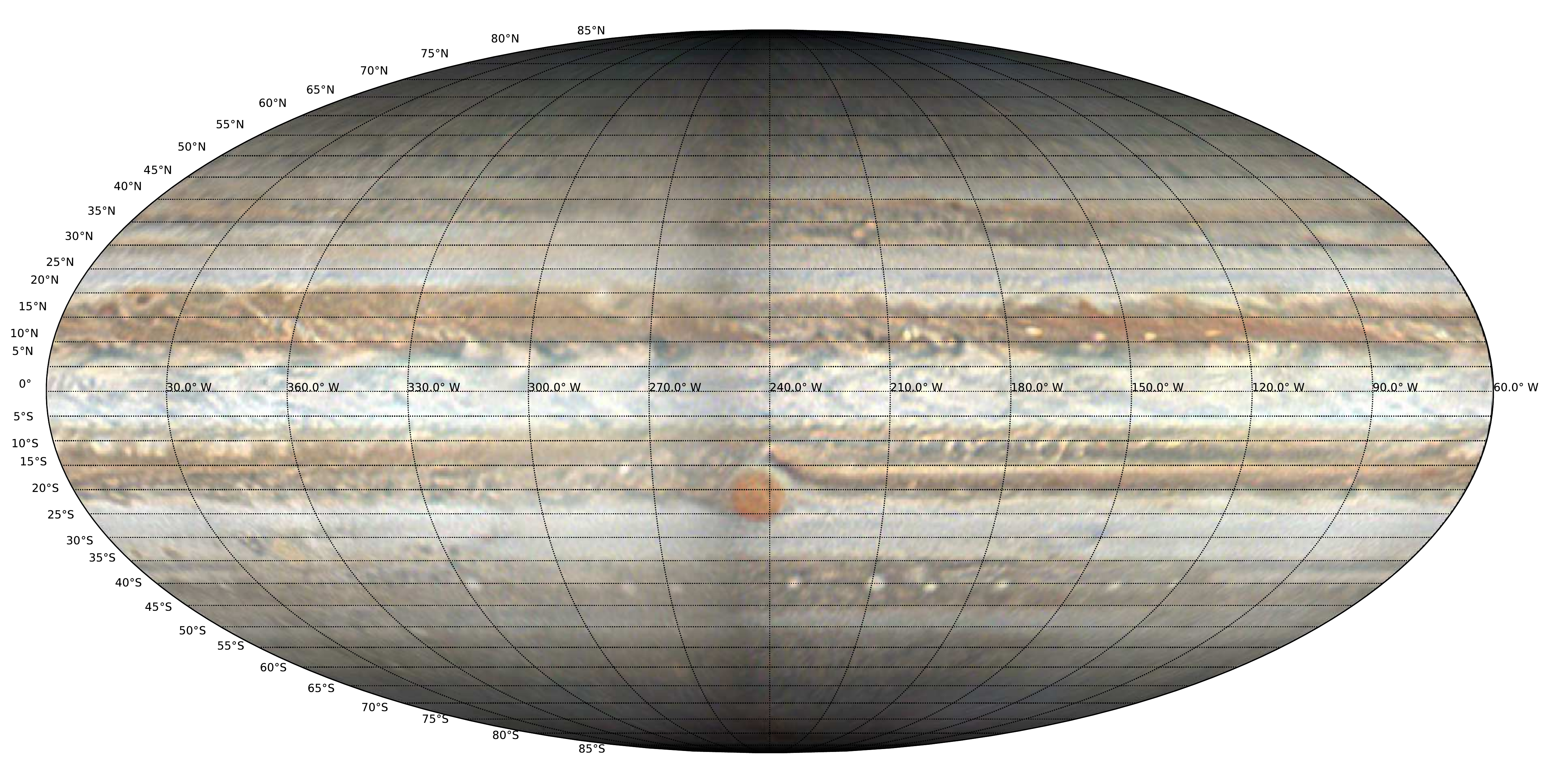}
\caption{\label{fig:map_visible_light} Planetocentric Mollweide projection of Jupiter in visible light. Longitudes are in system III. The mosaic was built from several pictures taken by \citet{Einaga2016} between 15 and 16 January 2016 using a 300 mm Newton telescope.}
\end{figure*}

\begin{figure*}[p]
\centering
\includegraphics[width=\textwidth]{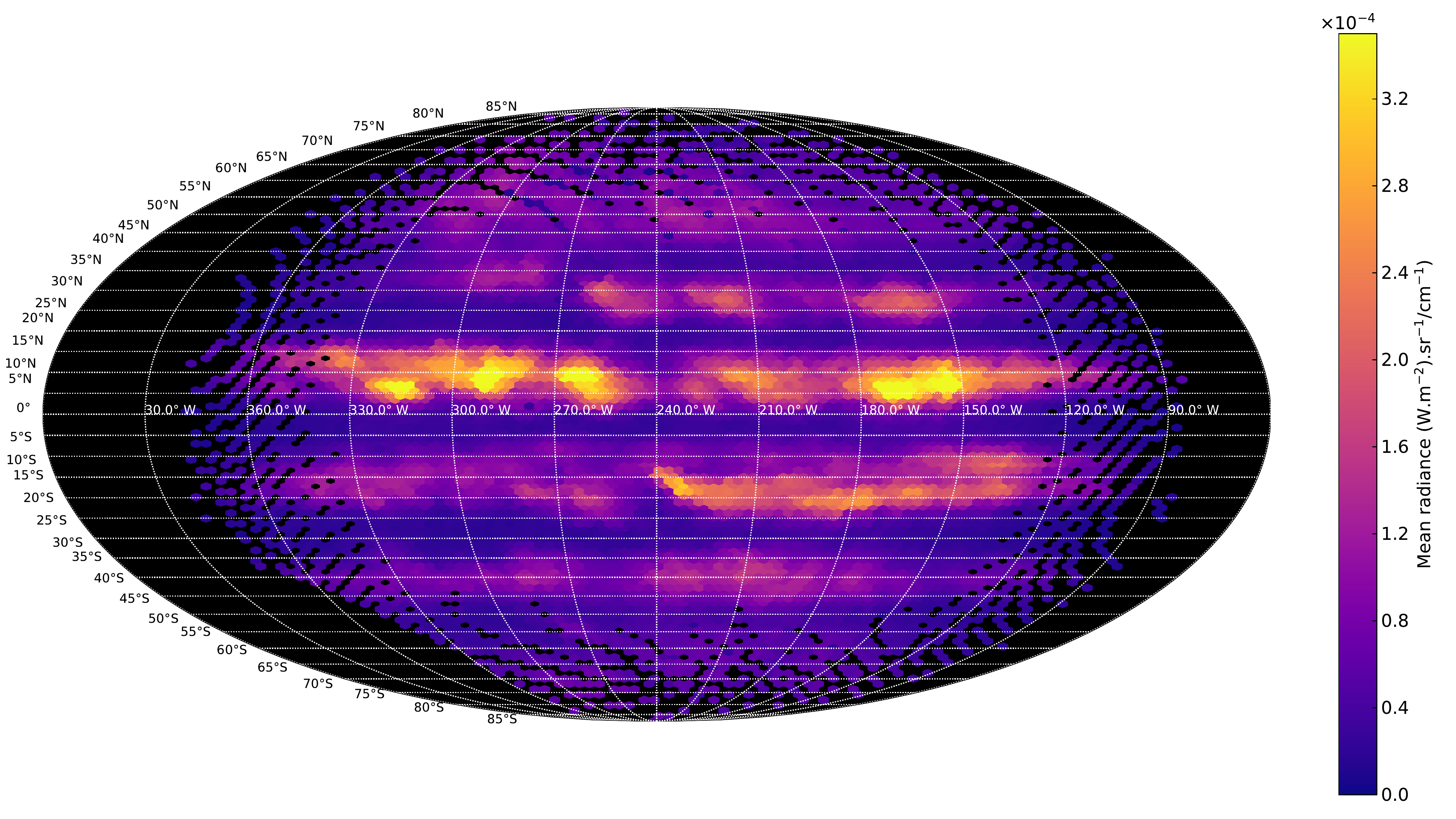}
\caption{\label{fig:map_radiances}Planetocentric Mollweide projection of the observed radiance of all our reduced IRTF/TEXES spectral cubes in the 1930--1943 cm$^{-1}$ wavenumber range. Longitudes are in system III. The Great Red Spot is located between latitudes 15--25$^\circ$S and longitudes 230--250$^\circ$W.}
\end{figure*}

\begin{table*}[t]
	\begin{center}
		\caption{\label{tab:spectral_cube_parameters}Reduced spectral cubes parameters, recorded on 16 January 2016}
		\begin{threeparttable}
		\begin{tabular}{c c c c c c c c}
			\hline
			Wavenumber range & Main & Observing time & Size & Longitude converage& Number of & $c_{H_2O}$\tnote{(1)} \\
			(cm$^{-1}$) & feature & (UT) & (px) & (sys. III $^\circ$W) & retrieved spectra & (mm) \\
			\hline
			\multirow{3}{*}{1930--1943} & \multirow{3}{*}{NH$_3$} & 09h36 &  $61\times95$ & 83--264 & 4489 & 1.5\\
			& & 11h10 &  $61\times95$ & 153--296 & 4472 & 2.0\\
			& & 12h50 &  $59\times96$ & 213--358 & 4511 & 2.5\\
			\hline
		\end{tabular}
		\begin{tablenotes}
		\item[(1)] Line-of-sight column density in the Earth's atmosphere. See section \ref{subsec:methods_transmission_earth}.
		\end{tablenotes}
		\end{threeparttable}
	\end{center}
\end{table*}

\begin{table*}[t]
	\begin{center}
		\caption{\label{tab:molecular_parameters}Molecular parameters and references}
		\begin{threeparttable}
		\begin{tabular}{l c c c}
			\hline
			\multirow{2}{*}{Molecule} & \multirow{2}{*}{VMR} & Line parameters & Broadening \\
			 & & ($\nu$, intensity and E$_\text{low}$) & parameters \\
			\hline
			H$_2$ & 0.863\tnote{(1)} & -- & -- \\
			He & 0.135\tnote{(1)} & -- & -- \\
			CH$_4$ & 1.81$\times$10$^{-3}$ & GEISA 2015 & GEISA 2015\\
			H$_2$O & \citet{Roos-Serote1998}\tnote{(2)} & GEISA 2015 & GEISA 2015\tnote{(3)} \\
			NH$_3$ & \citet{Sromovsky1998}\tnote{(4)} & GEISA 2015 & \citet{Brown1994} \\
			\hline
		\end{tabular}
		\begin{tablenotes}
			\item[(1)] From \citet{vonZahn1998}.
			\item[(2)] Vertical profile taken from cited article.
			\item[(3)] Pressure broadening at half width half maximum multiplied by 0.79 according to \citet{Langlois1994}.
			\item[(4)] We took the a priori of the vertical profile used in the article for all pressures lower than 0.8 bar. For greater pressures, we modified the value in some of our configurations (see section~\ref{subsec:methods_retrieval_method}).
		\end{tablenotes}
		\end{threeparttable}
	\end{center}
\end{table*}

\section{Observations and data reduction}
\label{sec:observations}
\subsection{Observations}
\label{subsec:observations_observations}
\subsubsection{IRTF}
We used three spectral cubes of Jupiter obtained on 16 January 2016 (UT) using Texas Echelon X-Echelle Spectrograph \citep[TEXES, see][]{Lacy2002} mounted on the InfraRed Telescope Facility (IRTF) at Mauna Kea, Hawaii. The data were acquired with the medium-resolution 1.4''$\times$45'' slit with the long axis aligned along celestial N/S. A visual approximation of the appearance of Jupiter at this date is shown in figure~\ref{fig:disk_visible_light}.

The slit was offset from Jupiter's center by 25'' west and stepped by 0.7'' increments east until the slit fell off the planet on the eastern limb.  Sky observations taken at the beginning and end of the scan were used to subtract the sky emission throughout the scan. The observations were flat fielded and flux calibrated using observations of a blackbody card placed in front of the instrument window at the beginning of the observations following the black-sky method described in \citet{Lacy2002}. This reduction is performed within the TEXES pipeline software along with wavelength calibration by using telluric lines within the bandpass observed and distortion corrections for all the optical effects within the spectrograph to return a fully reduced flux-calibrated 3-dimensional data cube, 2-d spatial and 1-d spectral. Then, the latitude and longitude of each pixel are determined. This is discussed in section~\ref{subsec:observations_latitudes_longitudes}.

The spectral cubes were transposed to make them look like the disk as observed from Earth, as shown for example in Figure~\ref{fig:disk_texes}. They cover the planet's disk in roughly 65$\times$95 pixels (depending on the cube), with a pixel-projected angular resolution of $\approx$0.7'' (the length of the scan steps) along the $x$ axis and $\approx$0.34'' along the $y$ axis (close to the diffraction limit of the telescope), with a spectral resolution of $\approx$0.15 cm$^{-1}$ in the 1930--1943 cm$^{-1}$ spectral range. This range permits us to probe the atmosphere in the 1--3 bar region via the strong NH$_3$ line at 1939 cm$^{-1}$, as shown later. 

The noise of the observed radiance at each wavenumber was taken as the mean of the standard deviation of the radiance of the off-disk pixels in the four 5$\times$5 pixels squares at each corner of the spectral cubes. This gives us a mean S/N ratio per spectral pixel of $\approx$10, with a maximum of $\approx$35. More details about the spectral cubes can be found in Table~\ref{tab:spectral_cube_parameters}.

It should be noted that there is known radiance calibration problems with TEXES, as highlighted by \citet{Fletcher2016} and \citet{Melin2017}. This may affect the absolute values, essentially of the cloud transmittance, but not the relative spatial variations that we observe.

\subsubsection{Visible light}
In order to compare our infrared observations with the visible aspect of Jupiter, we took a cylindrical map of Jupiter made by \citet{Einaga2016} from a mosaic of pictures taken between 15 and 16 January 2016 with a 300 mm Newton telescope and transformed it using Molleweide and orthographic projections. The results are displayed in figures~\ref{fig:disk_visible_light} and~\ref{fig:map_visible_light}.

\subsection{Latitudes and longitudes}
\label{subsec:observations_latitudes_longitudes}
After the pipeline processing, the data are pushed through a purpose-built IDL program where the user visibly matches the limb of the planet to an ellipse to locate the center of the planet.  Then using NAIF's ICY toolkit \citep{Acton1996} the program calculates the latitude and longitude of each pixel of the map in Jovian west longitude and planetocentric latitude.

One of the main challenges of using this method is that, with TEXES in the spectral range of our observations, the limbs are not well-defined. It is therefore hard to correctly place the ellipse. A few arc-seconds error on the placement of the center of the ellipse --- resulting on a few degrees error on latitudes and longitudes at the center of the disk --- or the size of the ellipse can result in errors reaching more than 10$^\circ$ on latitude and longitude near the limbs. This makes quantitative spatial comparisons between our work and others difficult. Hence, when spatially comparing our results with other works we favour a qualitative discussion.

\subsection{Doppler shift}
We corrected the Doppler shift of the observed spectra using a velocity map produced by the IDL mapping code, which takes into account the relative velocity of Jupiter with respect to Earth as well as Jupiter's rotation, so that each spectrum has its own Doppler-shift correction. To take this into account when adding the effect of the sky to the synthetic spectra (see section~\ref{subsec:methods_transmission_earth}), the sky is accordingly shifted in wavenumber. For information, the Doppler shift had a mean value of +0.17 cm$^{-1}$.

\section{Methods}
\label{sec:methods}
\subsection{Radiative transfer}
\label{subsec:methods_radiative_transfert}
The radiative transfer we use includes H$_2$--H$_2$ and H$_2$--He absorption, rovibrational bands from CH$_4$, H$_2$O and NH$_3$, as well as cloud-induced absorption, reflection and emission. The H$_2$--H$_2$ and H$_2$--He absorption were given by a subroutine originally written by A. Borysow and based on models by \citet{Borysow1985, Borysow1988}. The volume mixing ratios (VMR), line parameters (i.e. position, intensity and energy of the lower transition level), temperature and broadening coefficients references used for modelling these different gases are listed in Table~\ref{tab:molecular_parameters}. We used linewidths broadened by H$_2$ and He and their respective temperature dependence whenever available. For the line shape, we used a Voigt profile with a cut-off at 35 cm$^{-1}$ for all the molecules. 

Our initial a priori temperature profile comes from the measurement of the Galileo probe \citep{Seiff1998}. Our model atmosphere is divided up in 126 atmospheric layers from $10^{-7}$ to 20 bar, evenly distributed logarithmically.

Our model includes one monolayer grey (i.e. spectrally constant) cloud located at $0.8$ bar, with a lambertian reflectance ($I/F$) always equal to 0.15. This cloud is represented in our code by a scalar fixed to 1 above the cloud level and the cloud transmittance $t_c$ at and below the cloud level. This cloud model is inspired from the work of \citet{Giles2015, Giles2017}, who showed that the tropospheric cloud layer can be located between 1.2 and 0.8 bar, and that more refined cloud structures, as described by \citet{Atreya1999} for example (three multilayer clouds with a smooth transmittance gradient), have only a minor effect on the goodness of fit. For the spectral directional reflectance, we used the value retrieved by \citet{Drossart1998}, based on a comparison between 5-$\mu$m low-flux dayside and nightside spectra of Jupiter. However, we do not include multiple-diffusion or a deep cloud layer at 5 bar. These parameters play an important role only inside the zones \citep[see][for example]{Giles2015}, but in our data, the mean signal-to-noise ratio of these regions is systematically lower than 6, which is not high enough to retrieve valuable information. Therefore, our approach ignoring these parameters should have only a minor impact on our results.

The radiance is convolved to simulate the instrument function, which was approximated as a gaussian with a FWHM of $0.15$ cm$^{-1}$. 

Further details concerning the radiative transfer equations used can be found in the appendix.

\subsection{Transmission of the Earth's atmosphere}
\label{subsec:methods_transmission_earth}
To model the effect of Earth's atmospheric transmittance ("sky"), we used the code LBLRTM \citep{Clough2005} using a U.S. standard atmosphere and zenith angles and integrated column for telluric H$_2$O adjusted to reproduce the conditions of observation for each spectral cube.  Then, we multiplied our synthetic spectra and our retrieval derivatives (see section~\ref{subsec:methods_retrieval_method}) by the synthetic sky transmittances. In contrast with methods adopted by previous authors, where the observed radiance is divided by the sky transmittances \citep[i.e.][]{Bezard2002}, there is no need to remove some spectral ranges where the sky absorption is above an arbitrary limit. Concomitantly, the retrieval process is less sensitive to the radiances at the wavenumbers where the sky absorption is high.

We determined the zenith angles using the position and altitude of the telescope and the position of Jupiter in the sky at the date of observation coupled with ephemerides.

The value of the telluric H$_2$O column was derived, for each cube, using the jovian spectra with highest signal. We obtained the best fit with the synthetic spectrum generated with the line of sight column for telluric H$_2$O displayed in Table~\ref{tab:spectral_cube_parameters}.

\subsection{Retrieval method}
\label{subsec:methods_retrieval_method}
We used a classical optimal non-linear retrieval method \citep{Rodgers2000} to retrieve the abundance of the main molecular contributors at each atmospheric layer, as well as the cloud transmittance, while the abundance profiles of minor species as well as the temperature profile were kept constant. 

\begin{figure}[t]
\centering
\includegraphics[width=0.49\textwidth]{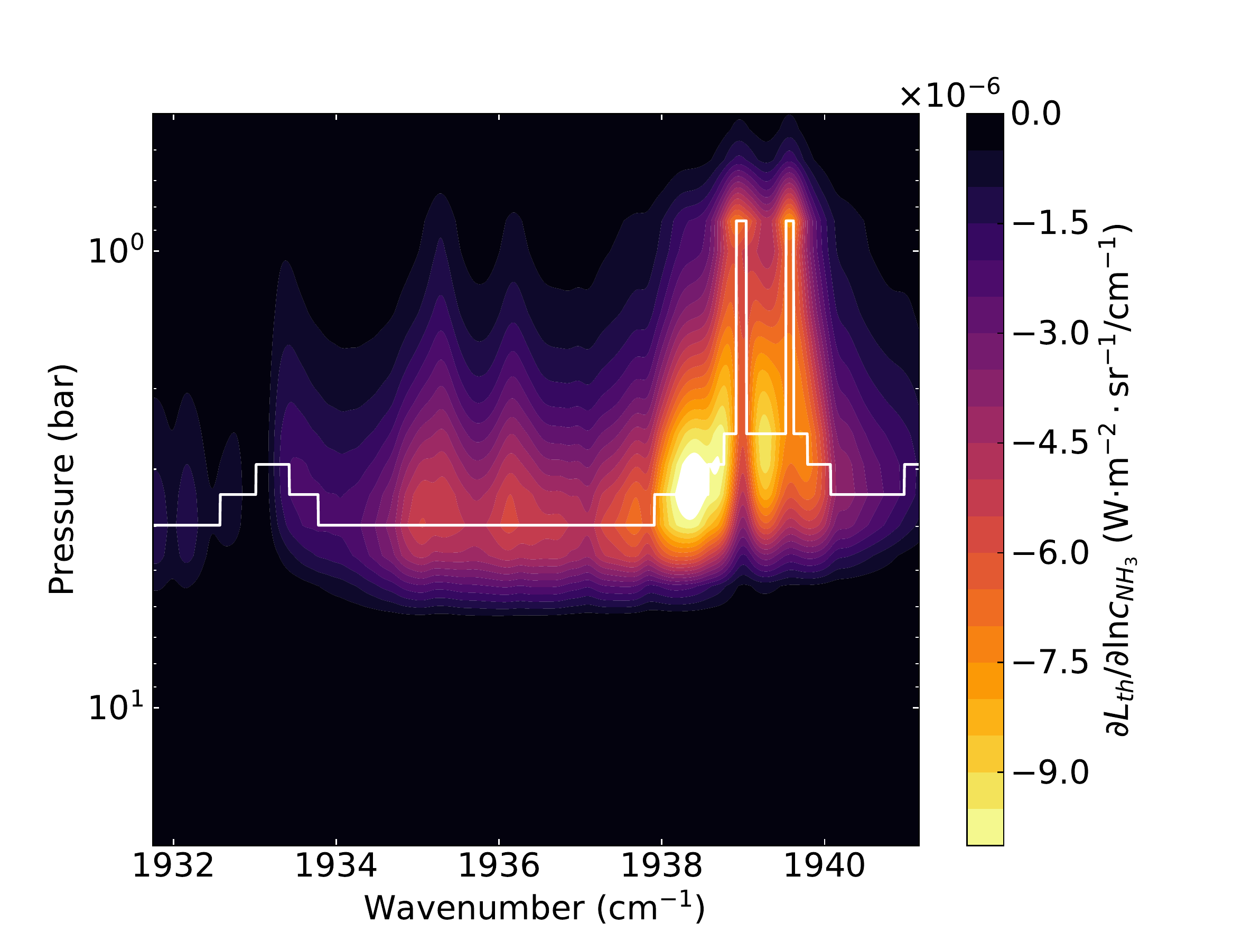}
\caption{\label{fig:nh3_kernel} Weighting function matrix for NH$_3$, with a cloud located at 0.8 bar. The cloud transmittance is 0.14 and its reflectance is 0.15. Solid white line: pressure of maximum sensitivity as a function of wavenumber.}
\end{figure}

This retrieval method can be described as follows. We can take the full state vector $x$ containing $k$ independent components, $x_1$, ..., $x_k$, representing in our case the abundance profiles and the cloud transmittance profile. It can be demonstrated (Rodgers, 2000, Eq. 4.29) that the best estimator $\hat{x}_j$ of a component $x_j$ of $x$ can be written as
\begin{equation}
\label{eq:rodgers200_best_estimator}
	\hat{x}_j = x_{aj} + S_{aj}K^T_j\left(\sum_{i=1}^{i=k} K_iS_{ai}K^T_i + S_\epsilon\right)^{-1}\Delta y
\end{equation}
With $x_{aj}$ the a priori estimate of $x_j$ (i.e. the a priori abundance profiles, listed in Table~\ref{tab:molecular_parameters}, and the cloud transmittance profile), $S_{a}$ the covariance matrix for the a priori parameters, which contains the uncertainties on each parameters, $K$ the weighting functions of the parameters, which contains the partial derivatives of the radiance with respect to the parameter (an example is displayed in figure~\ref{fig:nh3_kernel}), $S_\epsilon$ the measurement error diagonal covariance matrix, which is described in section~\ref{sec:observations}, and $\Delta y = y - y_a$ the difference between $y$, the observed radiance and $y_a$, the radiance calculated by our model using the a priori parameters.

The unweighted covariance matrix $\tilde{S}_{a}$, common to all the a priori parameters ($x_{a1}$, ..., $x_{ak}$) is given by:
\begin{equation}
\label{eq:unweighted_covariance_matrix}
	\tilde{S}_{a,xy} = \exp\left(-\log\left(p_x / p_y\right)^2 / 2v^2\right)
\end{equation}
With, in our case, $p_z$ the pressure at the atmospheric level $z$ and $v$ the vertical smoothing parameter, expressed in scale height. It determines the degree of smoothing applied to the solution. This matrix is then weighted by a factor $\sigma_j = w\text{~trace}(S_\epsilon) / \text{trace}(K_j\tilde{S}_{a}K_j^T)$, $w$ being the weight of the constraint on the departure from the a priori profile. In this study, we adopted $v = 0.75$ and $w = 0.1$. These values give us a fair goodness of fit and convergence speed, while preventing the abundance profile to oscillate. Note that for $w$, a wide range of values is possible (from 1 to 0.05), most of the time without significant changes in the results. If $w$ is taken too low or too high, the results might vary with the chosen a priori. Finally, the covariance matrix $S_{aj}$ for the a priori parameter $x_{aj}$ is given by:
\begin{equation}
\label{eq:covariance_matrix}
	S_{aj} = \sigma_j \tilde{S}_{a}
\end{equation}

Our algorithm follows these steps: (i) we calculate the radiance $y_a$ and its derivatives $K$, using our a priori profiles $x_{a}$. (ii) we estimate the goodness of fit of the modelled radiance $y_a$ on the observed radiance $y$ through the weighted sum of squared deviations per degrees of freedom (reduced $\chi^2$) function. (iii) we use our retrieval method described in Eq.~(\ref{eq:rodgers200_best_estimator}),
this gives us our new profiles $\hat{x}$. (iv) we update our a priori profiles  with our new profiles, so that $x_a = \hat{x}$. These four steps constitute one iteration. While the goodness of fit calculated in step 2 is improved by more than a threshold $\delta$, we continue to iterate, until the condition is fulfilled. For our runs, we choose a $\delta$ of $1 \%$. Typically, it takes about 10 iterations before the $\chi^2$ improvement drops below this threshold. We also limited the maximum number of iterations to 32 to speed up our retrievals. Increasing the maximum number of iterations does not significantly change our results. This limit was reached in less than 0.001$\%$ of the retrievals, and concerned mostly pixels with high cloud transmittance ($>$ 0.129). In the worst case, the $\chi^2$ was improved by 1$\%$ during the last iteration. We consider that, given the low improvement of the goodness of fit at the last iteration and the very limited number of spectra concerned, this limit in iteration number had no significant effect on our results.

To prevent the retrieval of being trapped within a $\chi^2$ local minima, we chose to make a first run retrieving only the cloud transmittance, while leaving the mixing ratios of the gases constant. In this case, we used a weight $w = 0.5$ to ensure a fast convergence. This permitted us to have an a priori value on the cloud transmittance for our second run very close to the solution value. In the second run, we retrieved all the abundance profiles as well as the cloud transmittance. It is this last run that gives us the spectra, profiles and maps that will be discussed in the following sections.

Clouds play a crucial role in our retrieval method. Because of the way clouds are taken into account (see the appendix) they can change drastically the sensitivity and therefore the retrieved profiles of our species. Indeed, the pressure level at which they are placed and the spectral directional reflectance value are critical. For example, the cloud-free maximum sensitivity at 5 $\mu$m of most of the species we study in this work is around 4 bar. If we add a cloud, the sensitivity at all the levels below the cloud level decreases due to its transmittance, while the sensitivity at and above the cloud level increases due to the solar reflected contribution. At some point, the level of maximum of sensitivity at a given wavenumber will switch from below the cloud level to above the cloud level. For NH$_3$ the effect is important even at a relatively high cloud transmittance of 0.1. The result is that, taking into account the effect of the smoothing matrix $S_a$, the maximum change in the retrieved abundance profile will be situated at lower pressure than in a cloud-free atmosphere, typically in the 1--3 bar range instead of the 3--4 bar expected.

For NH$_3$, we tested four configurations (A, B, C and D) in order to test the influence of the a priori and the classical view of a profile constant below a given altitude. (A) In the first configuration, we used the abundance profile from \citet{Sromovsky1998}, modified with a VMR of 200 ppmv for pressures greater than 0.8 bar, to represent the classical view on the abundance, but with a value at $\approx$ 10 bar close to what was found by \citet{Li2017} (B) In the second configuration, the a priori VMR was set to 300 ppmv for pressures greater than 0.8 bar, which is the value found by \citet{Sromovsky1998}. (C) In the third configuration, the a priori VMR was set to 400 ppmv for pressures greater than 0.8 bar, to have an a priori similar to the one used by \citet{Li2017}. (D) In the fourth configuration, we used the same a priori as in configuration B, but instead of allowing the profile to freely vary at each iteration, we impose that the profile must be constant at pressures greater than 2 bars, and at pressures comprised between 0.8 and 2 bars, taking as value the last retrieved VMR at 2 bars. The formulae of $K_{NH_3}$ and $S_{a,NH_3}$ are changed accordingly. This latter configuration represents a case where the NH$_3$ VMR is constant until its condensation at 0.8 bar. An example of a retrieval using these profiles is discussed in section~\ref{subsec:results_nh3}, the results are displayed on figures~\ref{fig:nh3_vmr_profiles_1110-38-36} and \ref{fig:nh3_vmr_profiles_zones}.

In the 1930--1943 cm$^{-1}$ spectral range, only H$_2$O --~and to a lesser extent, CH$_4$ and PH$_3$~-- play a significant role apart from NH$_3$. In this spectral range H$_2$O plays a role similar to the cloud transmittance, and retrieving its abundance profile following our methodology does not lead to significant improvement. Hence, we chose to not retrieve the H$_2$O abundance profile. Since we also assume the CH$_4$ VMR abundance profile to be constant, we chose to invert only the abundance profile of NH$_3$, while the abundance profiles of the other molecules were kept constant.

\subsection{Error handling}
The uncertainties on the parameters obtained by the retrieval method are given by the square root of the diagonal elements of the covariance matrix
\begin{equation}
\label{eq:error_covariance_matrix}
	\hat{S}_j = S_{aj} - G_jK_jS_{aj} + \mid G_j F\mid
\end{equation}
With $G_j = S_{aj}K^T_j\left(\sum_i K_iS_{ai}K^T_i + S_\epsilon\right)^{-1}$ the gain matrix and $F$ being 0 except on its diagonal where $F_{ii} = \Delta y_i$ (meaning no correlation between wavenumber). The first two terms represent the result of the sum of the smoothing error and the retrieval noise error (i.e. the propagation of the instrumental noise into the retrieved parameters). The last term represents the forward model error, which is due to the model imperfections. Note that the forward model error must be evaluated with the true parameters, rather than the retrieved parameters. However, at the end of all the iterations, the retrieved parameters are expected to be close to the true state, therefore the value used here should be a good approximation of the true forward model error.

\begin{figure*}[!p]
\centering
\includegraphics[width=\textwidth]{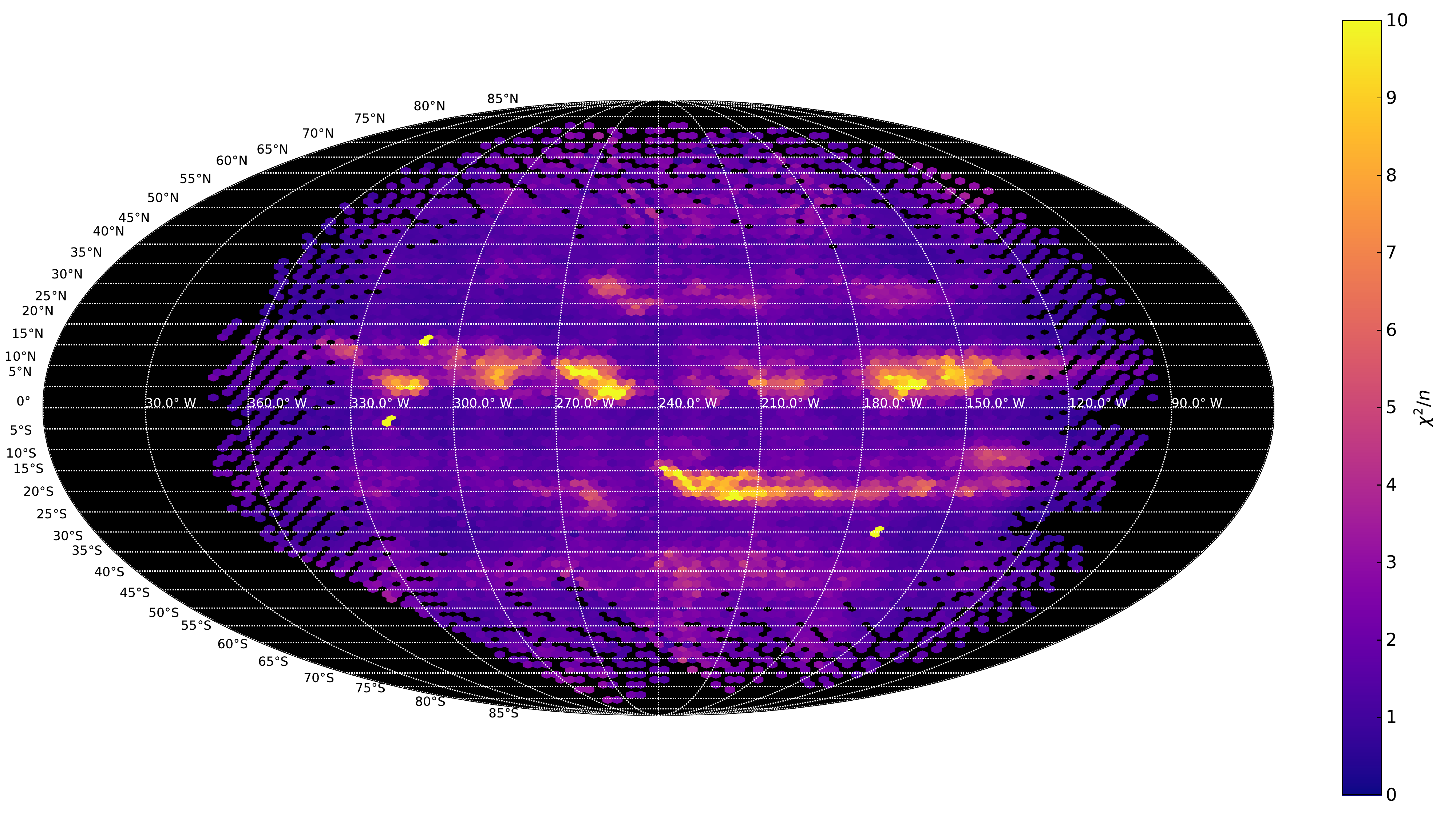}
\caption{\label{fig:map_chi2}Planetocentric Mollweide projection of the goodness of fit of all our reduced spectral cube in the 1930--1943 cm$^{-1}$ wavenumber range (configuration A). Longitudes are in system III.}
\end{figure*}

\begin{figure*}[!p]
\centering
\includegraphics[width=\textwidth]{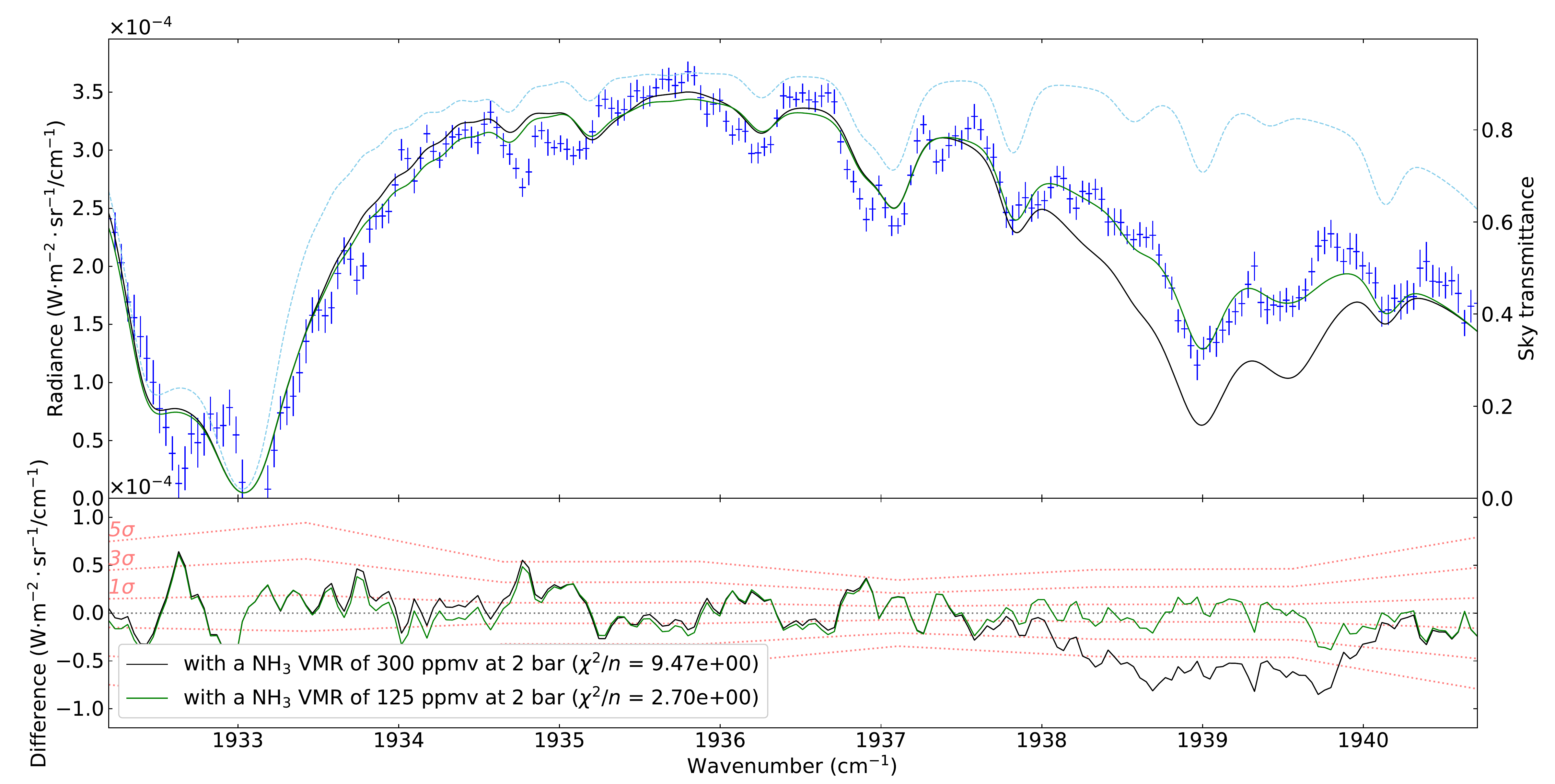}
\caption{\label{fig:spectrum_fit_example} Example of a fit on a relatively bright spectrum in the spectral cube taken at 11:10 UT situated at planetocentric latitude 8$^\circ$N and System III longitude 215$^\circ$W (configuration D). The wavelengths are given in the reference frame of Jupiter. The residuals are shown below. Dark blue: the observed spectrum, the errorbar represents the 1 sigma noise. Light blue: the convolved Earth's atmosphere transmittance, with a Doppler shift of -0.13 cm$^{-1}$. Green: our best fit, with a NH$_3$ VMR of 125 ppmv at 2 bar. Black: our a priori, with a NH$_3$ VMR of 300 ppmv at 2 bar. The NH$_3$ lines are around 1939.0 and 1939.5 cm$^{-1}$.}
\end{figure*}

\begin{figure*}[t]
\centering
\includegraphics[width=\textwidth]{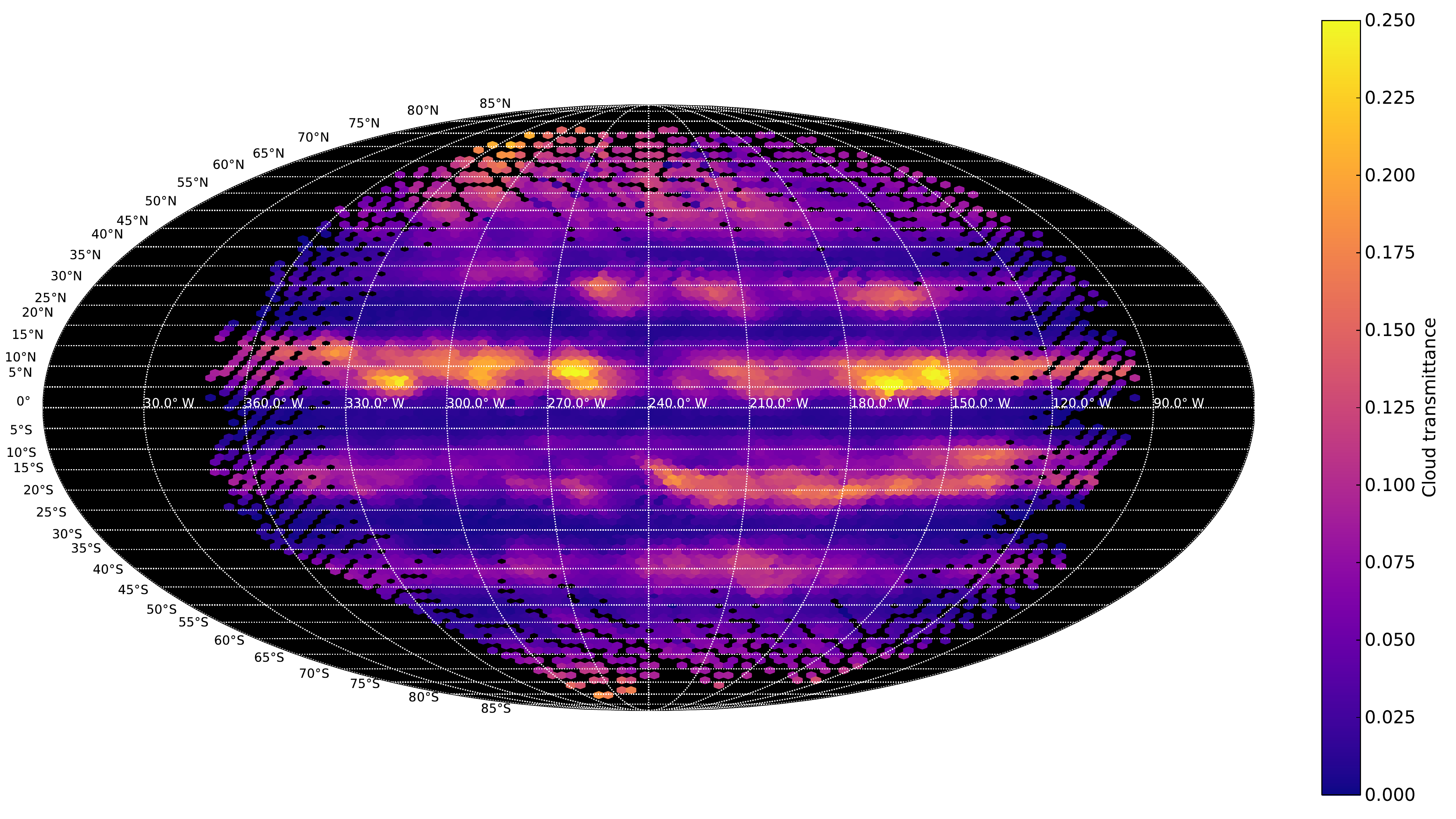}
\caption{\label{fig:map_cloud_transmittance} Planetocentric Mollweide projection of our retrieved 0.8-bar level cloud transmittance on all our reduced spectral cube in the 1930--1943 cm$^{-1}$ wavenumber range (configuration A). Longitudes are in system III.}
\end{figure*}

\section{Results and discussion}
\label{sec:results}
\subsection{Goodness of fit}
\label{subsec:results_chi2}
To calculate the goodness of fit of our retrievals we used the classical reduced RMS method ($\chi^2/n$, where $n$ is the number of free parameters, i.e. the number of samples in a spectrum). We obtained very similar goodness of fit for all our configurations, as shown in Table~\ref{tab:chi2_configurations}. Configurations A, B and C allow the NH$_3$ VMR profile to vary freely, hence the VMR profiles retrieved with these configurations can be difficult to physically explain. In contrast, the explanation for VMR profiles retrieved with configuration is straightforward: there is no source of NH$_3$ and the gas condenses into NH$_3$-ice clouds at 0.8 bar. More other, the goodness of fit obtained with this configuration is comparable to those of the other configurations. Therefore, we will discuss only the results derived from configuration D. A map of these RMS can be seen in figure~\ref{fig:map_chi2}. For configuration D, we obtained a mean  $\chi^2/n$ of 1.9 with a standard deviation of 1.3, and values lower than 4 for 95$\%$ of the retrieved spectra.

There is a large correlation between the mean radiance and the goodness of fit, due to the better SNR in high-flux spectra.
A comparison between a synthetic spectrum and an observed spectrum is displayed in figure~\ref{fig:spectrum_fit_example}.

These relatively high $\chi^2/n$ values are primarily explained by the presence in the majority of our observations of "spikes" and "dips" at roughly constant wavenumbers, of varying intensities and shapes, that we were unable to fit (there are some in figure~\ref{fig:spectrum_fit_example} at $\approx$ 1933.6, 1934.9, 1937.0 or 1939.9 cm$^{-1}$). We cannot definitively attribute those features to a specific instrument artifact, since its seems that the features follow Jupiter's band structure, but we strongly favour this explanation, for the following reasons. (i) In the spectra where these features seem insignificant, we are able to obtain reasonably good fits ($\chi^2$/n $\approx$ 1). (ii) Our model was tested and validated on a spectrum of Jupiter in the same spectral range, already analysed in \citet{Bezard2002}, so our model should not be the main issue. (iii) These features do not correspond to lines of any simple constituent, though some of them seem to be correlated with the telluric absorption. (iv) Previous works on TEXES in the same spectral range, such as \citet{Fletcher2016} did not mention such issues.

\begin{table}[t]
	\begin{center}
		\caption{\label{tab:chi2_configurations}Goodness of fit for each configuration for all the retrieved spectra}
		\begin{threeparttable}
		\begin{tabular}{c c c c}
			\hline
			Configuration & NH$_3$ a priori "deep" & \multirow{2}{*}{$\overline{\chi^2/n}$} & \multirow{2}{*}{$\sigma_{\chi^2/n}$} \\
			name & abundance (ppmv) && \\
			\hline
			A & 200\tnote{(1)} & 1.852 & 1.308 \\
			B & 300\tnote{(1)} & 1.850 & 1.305 \\
			C & 400\tnote{(1)} & 1.869 & 1.332  \\
			D & 300\tnote{(2)} & 1.907 & 1.307  \\
			\hline
		\end{tabular}
		\begin{tablenotes}
		\item[(1)] For pressures greater than 0.7 bar.
		\item[(2)] Same than (1) for the first iteration. For iteration $n$: pressures greater than 1.8 bar and between 0.8 and 1.8 bar set to the abundance value at 1.8 bar retrieved at iteration $n - 1$.
		\end{tablenotes}
		\end{threeparttable}
	\end{center}
\end{table}

Our uncertainties take into account the capacity of our model to fit the data, and the NH$_3$ abundance of the regions exempt of those features is consistent with the regions where they are more intense, so we stay confident in our results.

\subsection{Cloud transmittance}
\label{subsec:results_cloud_transmittance}
It should be kept in mind that our methodology assumes that the main modulator of flux is cloud transmittance rather than absorption by gases. Therefore, it is not surprising that our retrieved cloud transmittances are strongly correlated with the mean radiance map. Once again, the results are very similar among configuration A, B, C and D, so we will discuss only configuration D.

It can be seen from figure~\ref{fig:map_cloud_transmittance} that the zones are very cloudy, contrary to the belts. The northern belts are organised in patches of relatively cloud-free regions, which correspond to dark, bluish regions south of the belts, at the interface with the white zones --- the so called "hotspots". It can also be seen that the globally thinnest cloud patches are located in the North Equatorial Belt (NEB), with a cloud transmittance greater than 0.2. The region at $\approx$ 315$^\circ$W in the NEB is a good example. It should be noted that the whitest regions of the northern belts in the visible image (figure~\ref{fig:map_visible_light}) often correspond to the regions with the thickest clouds. This can be seen for example near 240$^\circ$W in the NEB or near 210$^\circ$W in the North Temperate Belt (NTB).

In the South Equatorial Belt (SEB), the Great Red Spot (GRS), situated at longitude 240$^\circ$W, seems to perturb a line of thin clouds (in the visible (c.f. figure~\ref{fig:map_visible_light}), the dark brown line between 15 and 20$^\circ$S) from its east side, so that the entire band west of the GRS is covered by thick, light brown clouds that get more and more transparent westward. This feature has been observed for example by \citet{Giles2015} with 2001 Cassini's VIMS spectral cubes. It can be explained by turbulence caused by the GRS forming clouds. South-west of the GRS, there is another line of thinner clouds at 20$^\circ$S, between 240 and 270$^\circ$W, which correspond to a dark bluish region in the visible image. This region seems to circle around the GRS, similarly to the dark line east of the GRS. Still in the SEB, between longitude 120$^\circ$W and 150$^\circ$W, it seems that there are two dark brown, transparency filaments, one at the north and one at the south of the belt, separated by a thin line of white, thicker clouds. The southern one reaches the GRS, while the clouds of the northern one get thicker approaching the GRS.

In the South Temperate Belt (STB), the distribution seems to be simpler. The clouds get thinner at the center of the belt, and thicker at its borders. The white ovals that can be seen in the visible at latitude $\approx$ 40$^\circ$S do not seem to have an infrared cloud counterpart, but it might be because we do not have a high enough spatial resolution.

Quantitatively, these results are consistent with cloud opacities at 5 $\mu$m retrieved in other works, such as \citet{Bezard2002}, who found a transmittance of $\approx$ 0.45 inside hotspots, or \citet{Irwin2001}, who find transmittances between $\approx$ 0.20 and 0.30 at 2 bar in bright regions of the atmosphere. It is also consistent with precise cloud retrievals, such as by \citet{Wong2004}, who found that a compact grey cloud of transmittance 0.14 is needed to fit their observation in the NEB. 

\subsection{NH$_3$}
\label{subsec:results_nh3}
\begin{figure}[t]
\centering
\includegraphics[width=0.49\textwidth]{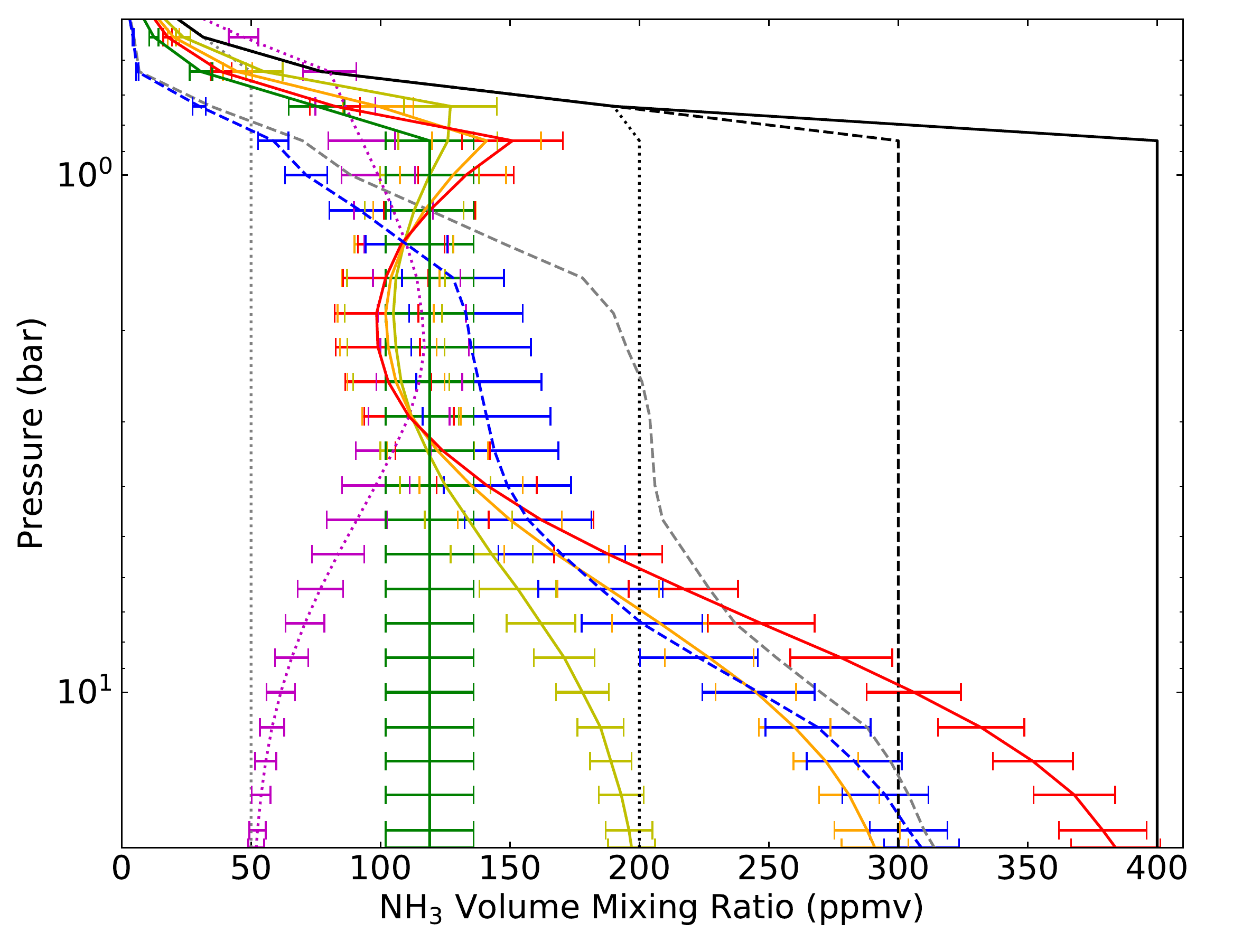}
\caption{\label{fig:nh3_vmr_profiles_1110-38-36}Retrieved NH$_3$ abundance profiles for the spectral cube taken at 11:10 UT situated at planetocentric latitude 8$^\circ$N and System III longitude 215$^\circ$W (see figure~\ref{fig:spectrum_fit_example}). Dotted black: configuration A a priori abundance profile. Dashed black: configuration B and first iteration of configuration D a priori abundance profile. Solid black: configuration C a priori abundance profile. Yellow, orange, red and green: retrieved abundance profile from respectively configuration A, B, C and D. Purple: retrieved abundance profile using the dotted grey a priori. Blue: retrieved abundance profile using the dashed grey a priori, meant to be close to what was retrieved by MWR at latitude 8$^{\circ}$N}.
\end{figure}

\begin{figure*}[p]
\centering
\vspace{-15pt}
\includegraphics[width=\textwidth]{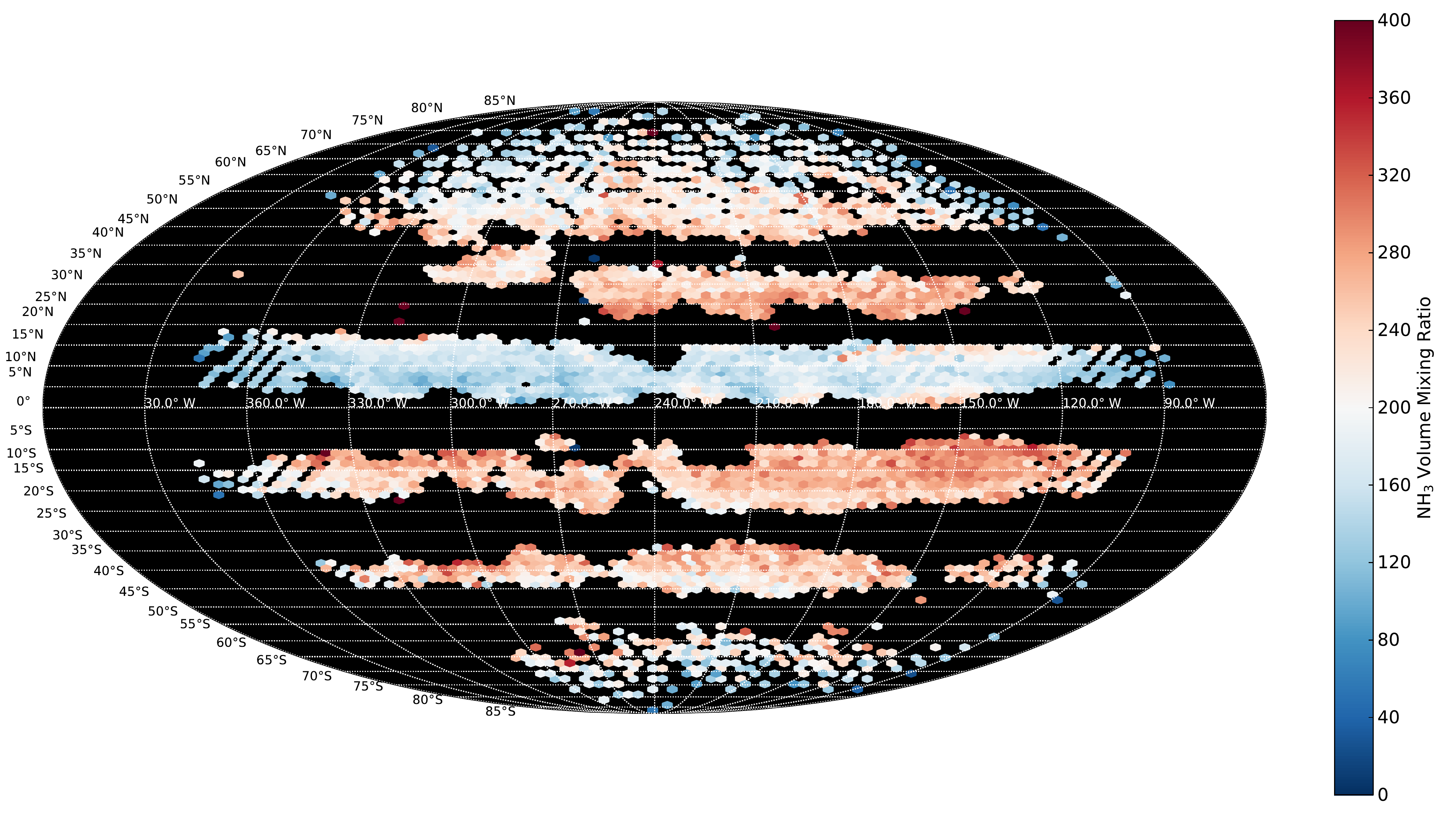}
\caption{\label{fig:map_nh3_vmr_2bar} Planetocentric Mollweide projection of our retrieved abundance of NH$_3$ at 2 bar (configuration D). Longitudes are in system III. All the points with a cloud transmittance lower than 0.05 has been removed.}
\end{figure*}

\begin{figure*}[p]
\centering
\includegraphics[width=\textwidth]{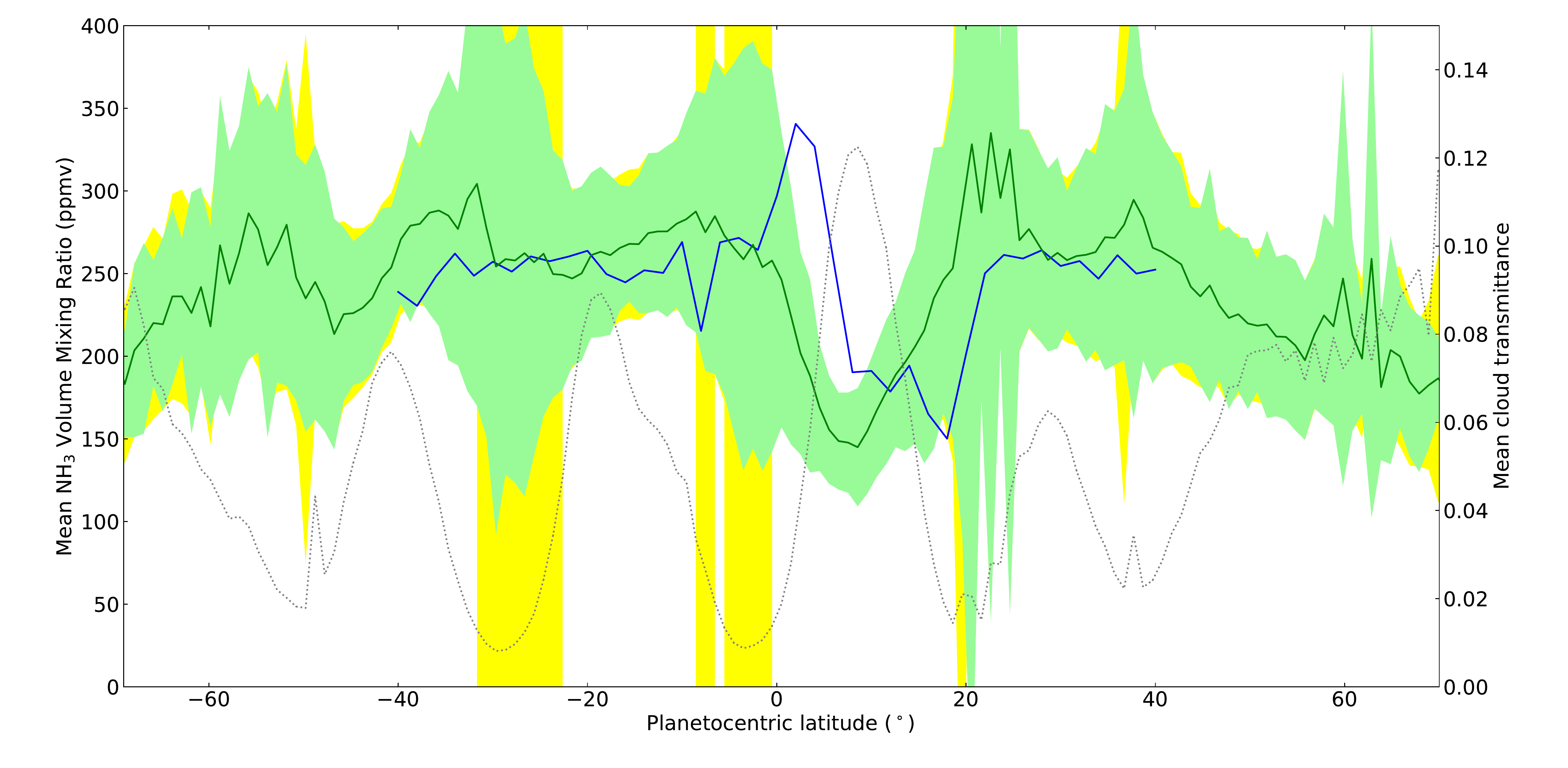}
\caption{\label{fig:nh3_vmr_latitudes} Zonally-averaged NH$_3$ abundance as a function of latitude. Green line: mean retrieved abundance of NH$_3$ at 2 bar over all our data. Green zone: standard deviation of our retrieved NH$_3$ abundance in 1$^\circ$ bins. Yellow zone: mean uncertainty (per spectrum) on the NH$_3$ abundance derived from the retrievals over the NH$_3$ abundance at 2 bar. Grey dotted line: mean retrieved cloud transmittance over all our data. Blue line: NH$_3$ abundance at 2 bar as measured by MWR during Juno's perijove 1 (PJ1); data obtained with the courtesy of C. Li, first published in \citet{Bolton2017}.}
\end{figure*}

The Jupiter spectra we have can be separated into two groups. One group is constituted of high-flux, high SNR spectra, which gives a reasonably low relative uncertainty on the retrieved NH$_3$ abundance. The other group, in contrast, is constituted of low-flux, low-SNR spectra, which gives a high uncertainty on the retrieved NH$_3$ abundance. One way to separate these groups is to use the retrieved cloud transmittance. Indeed, the retrieved cloud transmittance is highly correlated with the mean radiance --- and therefore, the flux --- of the spectra, and is easy to manipulate. If we set a cloud transmittance threshold of 0.05 to separate the two groups, it appears that all the spectra below this threshold have a low SNR ($\approx$ 4 or less) and are located exclusively in the zones. Hence, for now we will refer as  "zones" all the points with a retrieved cloud transmittance lower than 0.05, and as "belts" all the other points.

\subsubsection{Belts}
\label{subsubsec:results_nh3_belts}
An example of an abundance profile retrieved by each configuration is displayed in figure~\ref{fig:nh3_vmr_profiles_1110-38-36}. We used the same spectrum as for figure~\ref{fig:spectrum_fit_example}.
In this figure~\ref{fig:nh3_vmr_profiles_1110-38-36}, we can see that we obtain very similar results in the 1--3 bar region --- where our maximum of sensitivity lies --- with all our configurations, while anywhere else the results are very a priori-dependent and therefore not meaningful. Still in this figure, we also show the profiles retrieved using an a priori close to that retrieved by MWR \citep{Bolton2017} inside of the NEB and an a priori with an abundance set to 50 ppmv below 0.6 bar. The solution profiles are close to the other ones in the 1--3 bar range and confirms that, outside of this sensitivity region, no reliable information is available. We will therefore hereafter only show the retrievals for configuration D (figures~\ref{fig:map_nh3_vmr_2bar} and \ref{fig:nh3_vmr_latitudes}). 

In figure~\ref{fig:map_nh3_vmr_2bar}, we removed all the points in the zones. 
The mean uncertainty on the NH$_3$ VMR at 2 bar, outside of the zones, is $\approx$ 20$\%$. In this figure we observe a large depletion of NH$_3$ in the middle of the NEB, with a volume mixing ratio at 2 bar lower than 200 ppmv and going down to 60 ppmv, between planetocentric latitudes $\approx$ 0--17$^\circ$N, over all our longitude coverage (system III 90--360$^\circ$W). This depletion seems to be less significant at longitudes 225--240$^\circ$W, where thicker clouds are present.

The STB and the SEB are enriched compared to the NEB ($\approx$ 250 ppmv), and correspond to darker regions in the visible. The NTB is similarly enriched, except between 270--300$^\circ$W, where the NH$_3$ abundance is lower than 200 ppmv. Globally, the southern belts seem to be have a decreasing NH$_3$ abundance southward --- from $\approx$ 250 ppmv north of the belts to $\approx$ 200 ppmv south of the belts --- while it is the opposite for the NTB. In the SEB, the NH$_3$ abundance seems to be lower around the GRS at 17$^\circ$S, 230$^\circ$W ($\approx$ 200 ppmv), than in most of the belt ($\approx$ 300 ppmv). A filament of depleted NH$_3$ appears north of the SEB east of longitude $\approx$ 150$^\circ$W, correlated to the filament of thinner cloud observed in figure~\ref{fig:map_cloud_transmittance}. westward of the GRS, the NH$_3$ abundance seems to decreases with longitude.

Poleward of latitudes $\approx$ 45$^\circ$N, there seems to be a decreasing NH$_3$ abundance towards the Pole, starting at $\approx$ 240 ppmv and going down to $\approx$ 160 ppmv. We have only a few points poleward of latitude $\approx$ 50$^\circ$S, but the behaviour seems to resemble what we observed in the northern hemisphere.

Globally, the NH$_3$ abundance shows little variation with longitude in the temperate belts or near the poles, but the smallest features, like the "ovals", may not be resolved. In the SEB, the GRS seems to have a major influence, separating two very distinct behaviours we discussed earlier. In the NEB, the latitudinal width of the depletion seems to vary with longitude: 5$^\circ$ wide at longitudes 150$^\circ$W, 200$^\circ$W, 270$^\circ$W, almost disappearing at 235$^\circ$W, while nearly 10$^\circ$ wide in most of the belt. Strong longitudinal variations should be observed by MWR in the equatorial belts as it probes differents over different perijoves.

In figure~\ref{fig:nh3_vmr_latitudes}, we show the NH$_3$ mole fraction at 2 bar and the cloud transmittance longitudinally averaged over 1$^\circ$ latitude bins, and compare them with MWR retrieval \citep{Bolton2017}. As expected, there is a huge uncertainty on our results in the zones, and the mean abundance retrieved here should be taken cautiously. We also observe the same north/south NH$_3$ abundance "slope" in the belts (at the exception of the NEB) and poleward of latitude 45$^\circ$ that we observed in figure~\ref{fig:map_nh3_vmr_2bar}. The NEB depletion is also clearly visible. Globally, we obtain a mean NH$_3$ abundance close to what was found by MWR, but we do not see evidence of the "plume" detected by MWR in the EZ. This discrepancy can be explained both by our latitude uncertainty (discussed in section~\ref{subsec:observations_latitudes_longitudes}),  which can reach $\approx$ 5$^\circ$ at the equator, and by our abundance uncertainties in this region, which is greater than $\pm$ 300 ppmv. In summary, our results are in good agreement with MWR measurements. Our values of NH$_3$ VMR in the NEB are also consistent with the preliminary retrieval from JIRAM observations North of two hotspots \citep{Grassi2017}.

\subsubsection{Zones}
\label{subsubsec:results_nh3_zones}
\begin{figure}[t]
\centering
\includegraphics[width=0.49\textwidth]{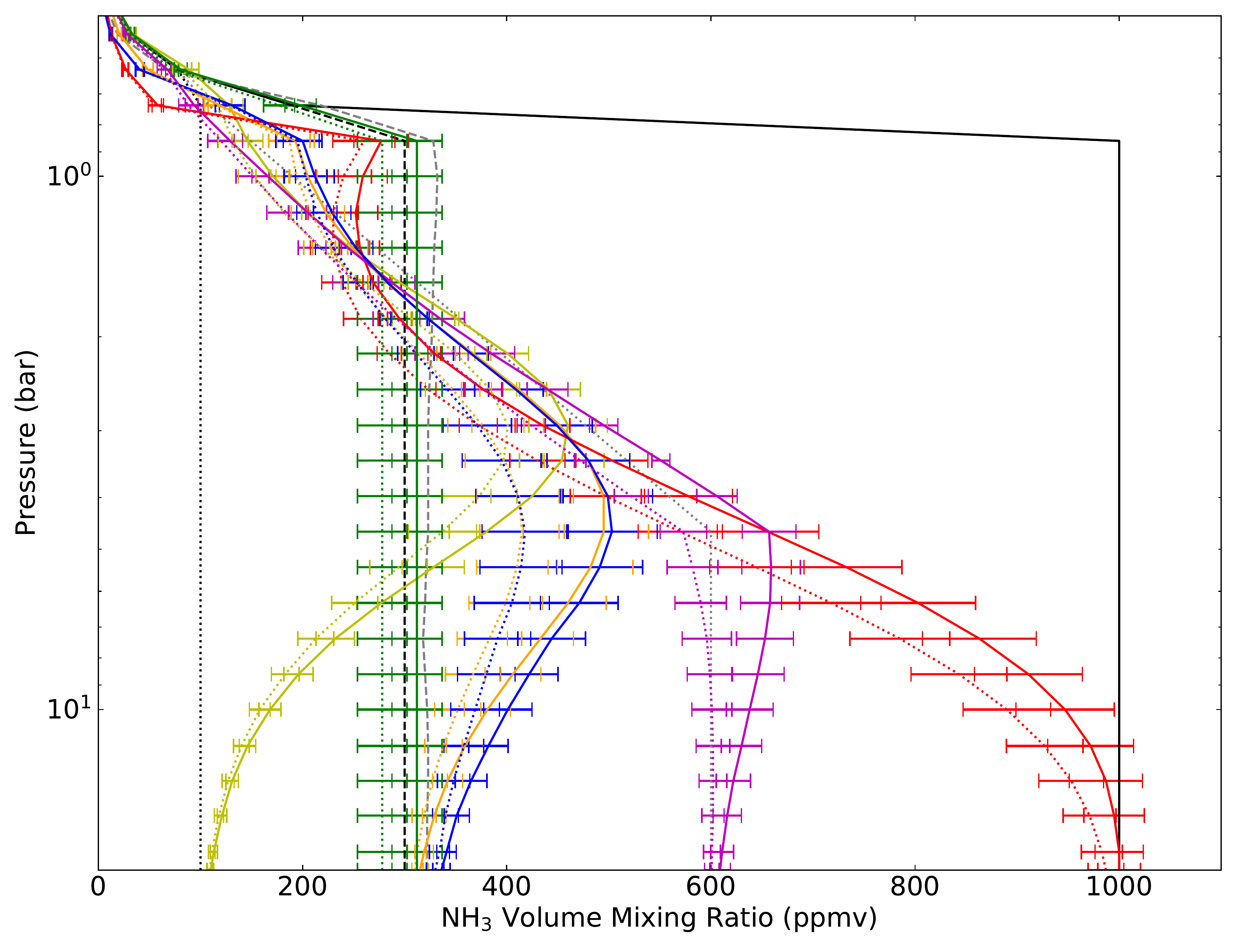}
\caption{\label{fig:nh3_vmr_profiles_zones}Retrieved NH$_3$ abundance profiles for the average of the spectra of all the zones (solid coloured curves) and for the average of the spectra of the EZ only (dotted coloured curves). Yellow, orange, red: retrieved NH$_3$ abundance profile for respectively the dotted black, dashed black and solid black a priori profiles. Green: retrieval similar to configuration D, for the dashed black a priori profile. Purple: retrieved NH$_3$ abundance profile for the dotted grey a priori profile, with a slope from 100 ppmv at 1 bar to 600 ppmv at 4 bar. Blue: retrieved NH$_3$ abundance profile for the dashed grey a priori profile, meant to be close to what was retrieved by MWR in the EZ.}
\end{figure}

\begin{figure*}[p]
\centering
\subfloat[\label{subfig:spectrum_fit_zones_spectrum_fit_zones}]{\includegraphics[width=\textwidth]{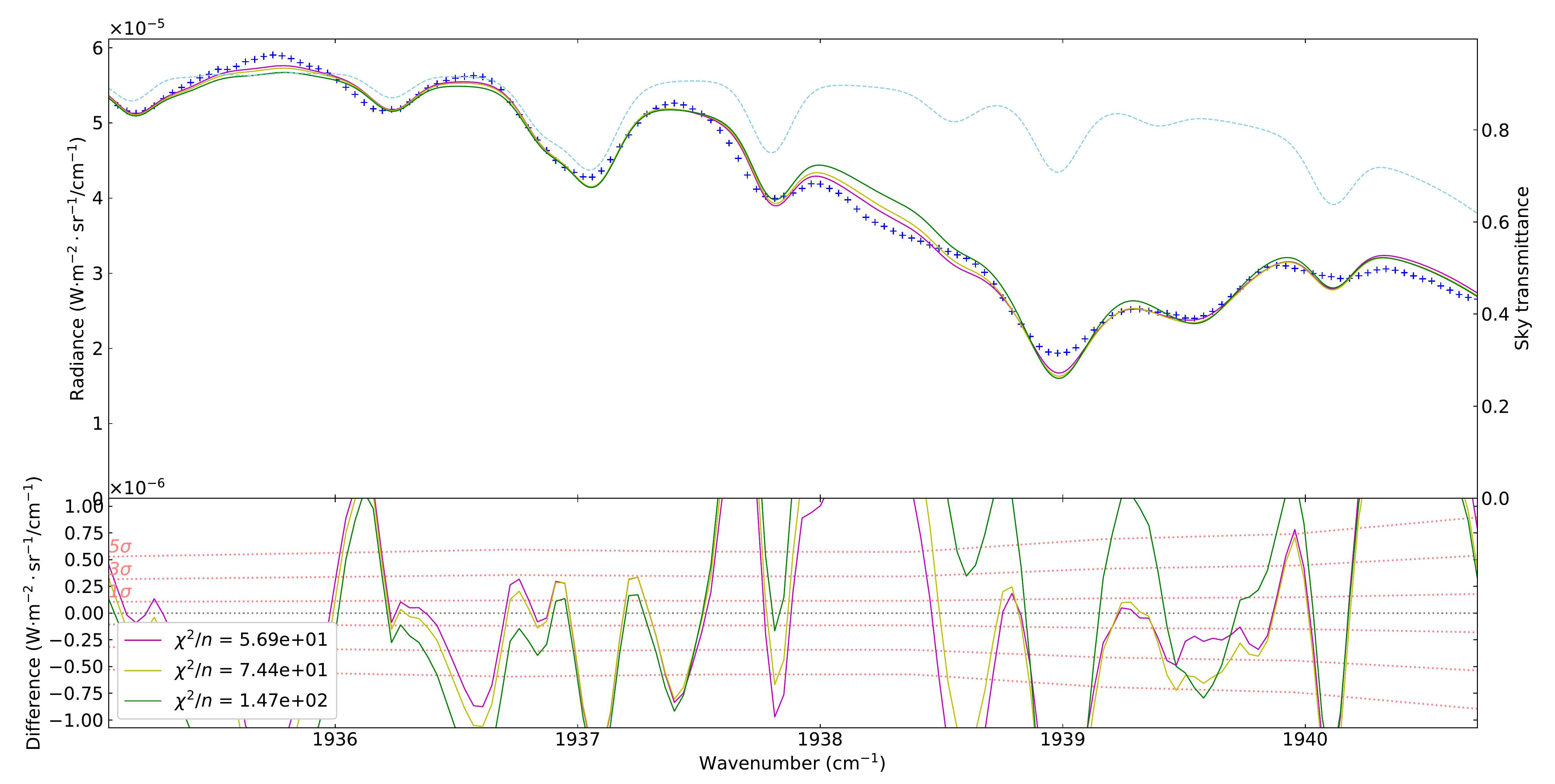}}
\\
\subfloat[\label{subfig:spectrum_fit_zones_spectrum_fit_ez}]{\includegraphics[width=\textwidth]{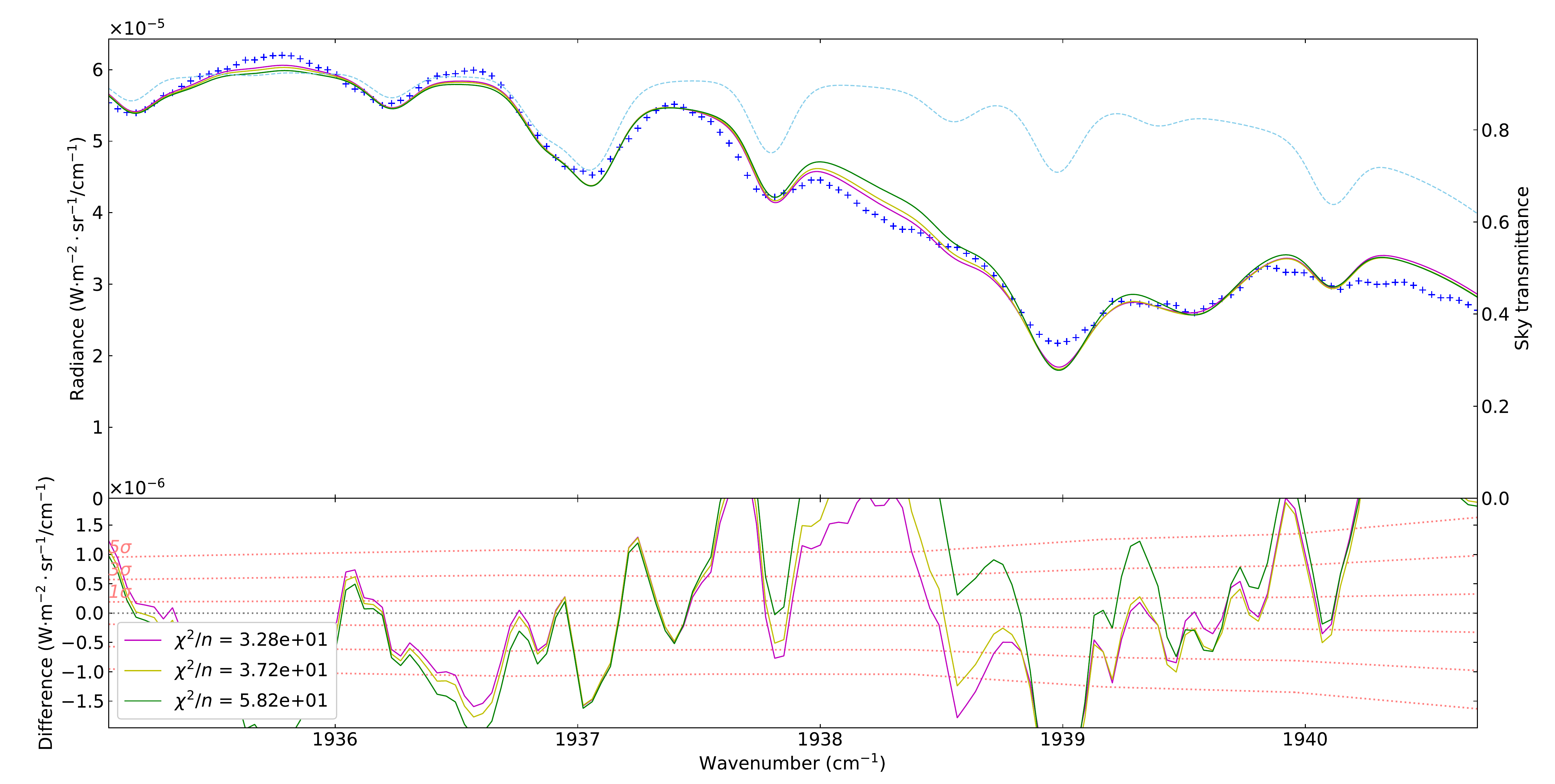}}
\caption{\label{fig:spectrum_fit_zones} Fit of the average of the spectra in all the zones (top) and of the average of the spectra in the EZ only (bottom). The wavelengths are given in the reference frame of Jupiter. The residuals are shown below. Dark blue: the average of the spectra, the errorbar represents the 1 sigma noise, divided by the square root of the number of the averaged spectra. Light blue: the convolved Earth's atmosphere transmittance, with a Doppler shift of -0.15 cm$^{-1}$, which is the mean of the Doppler shift of all the spectra in the zones. Yellow: our best fit using a NH$_3$ abundance profile a priori of 100 ppmv below 0.8 bar. Purple: our best fit using a NH$_3$ abundance profile a priori where the NH$_3$ is increasing with depth from 100 ppmv at 1 bar to 600 ppmv at 4 bar (see figure~\ref{fig:nh3_vmr_profiles_zones}).}
\end{figure*}

As mentioned above in this section, the results we obtain in the zones are much less reliable than in the belts, due to the low SNR of the spectra in these regions. 

In an attempt to enhance the SNR of the zones, we averaged all the spectra in these regions, as well as their corresponding Doppler shift and noise. Doing this allows us, in first approximation, to divide the noise by the square root of the number of spectra averaged. However, this approach has some limits. For example, the Doppler shift, the viewing angle, the solar incidence angle vary with the spectra, while the telluric absorption varies with the spectral cube, and these may not be well taken into account in the averaging. Hence, the results we obtain by this methodology should be taken particularly cautiously.

With the above-mentioned methodology, we obtained two spectra. (i) In the first one, we averaged all the spectra in the zones, (ii) while in the second one, we averaged the spectra located in the zones between latitudes 15 $^\circ$S and 5 $^\circ$N, corresponding to the EZ. 

In figure~\ref{fig:nh3_vmr_profiles_zones}, we display the retrieved NH$_3$ abundance profiles of the two spectra for various a priori. We can see that our domain of sensitivity remains roughly the same compared to what we obtain in the belts: between 1 and 3 bar. There is no significant differences between the behaviour of the retrieved profiles for the EZ (dotted curves) and for all the zones considered together (solid curves) in the 1--3 bar region. However, the behaviour of the NH$_3$ abundance profiles is very different from what we obtained in section~\ref{subsubsec:results_nh3_belts} (figure~\ref{fig:nh3_vmr_profiles_1110-38-36}). Instead of what can be interpreted as constant-with-depth abundance profiles in the 1--3 bar range, we obtain a NH$_3$ abundance that seems to increase with depth, from $\approx$ 100 ppmv at 1 bar to $\approx$ 500 ppmv at 3 bar. This is the opposite of what was found by MWR \citep{Bolton2017} both in the zones --- where the NH$_3$ abundance is observed to decrease with depth from $\approx$ 300 ppmv at 0.7 bar to $\approx$ 200 ppmv at 7 bar --- and in the EZ --- where the NH$_3$ abundance is observed to remain constant at $\approx$ 400 ppmv. What we obtain is closer to what was obtained in \citet{Giles2017}, with a NH$_3$ abundance increasing from $\approx$ 10 ppmv at 1 bar to $\approx$ 500 ppmv at 3 bar (or from 30 to 90 ppmv, depending on the cloud model used). These differences in retrieved values could be explained by the absence of light scattering in our model.

In figure~\ref{fig:spectrum_fit_zones}, we display our two averaged spectra and two of the fits we obtained. We voluntarily removed the 1932--1935 cm$^{-1}$ region which is dominated by telluric absorption. We can see that we obtain a relatively good fit, given the small noise considered. We can also see that the different fits are relatively similar despite their different initial a priori profiles.

\subsection{NH$_3$ VMR and cloud transmittance correlation}
\begin{figure*}[p]
\centering
\vspace{-15pt}
\subfloat[\label{subfig:nh3_vmr-cloud_transmittance_093600}]{\includegraphics[width=0.32\textwidth]{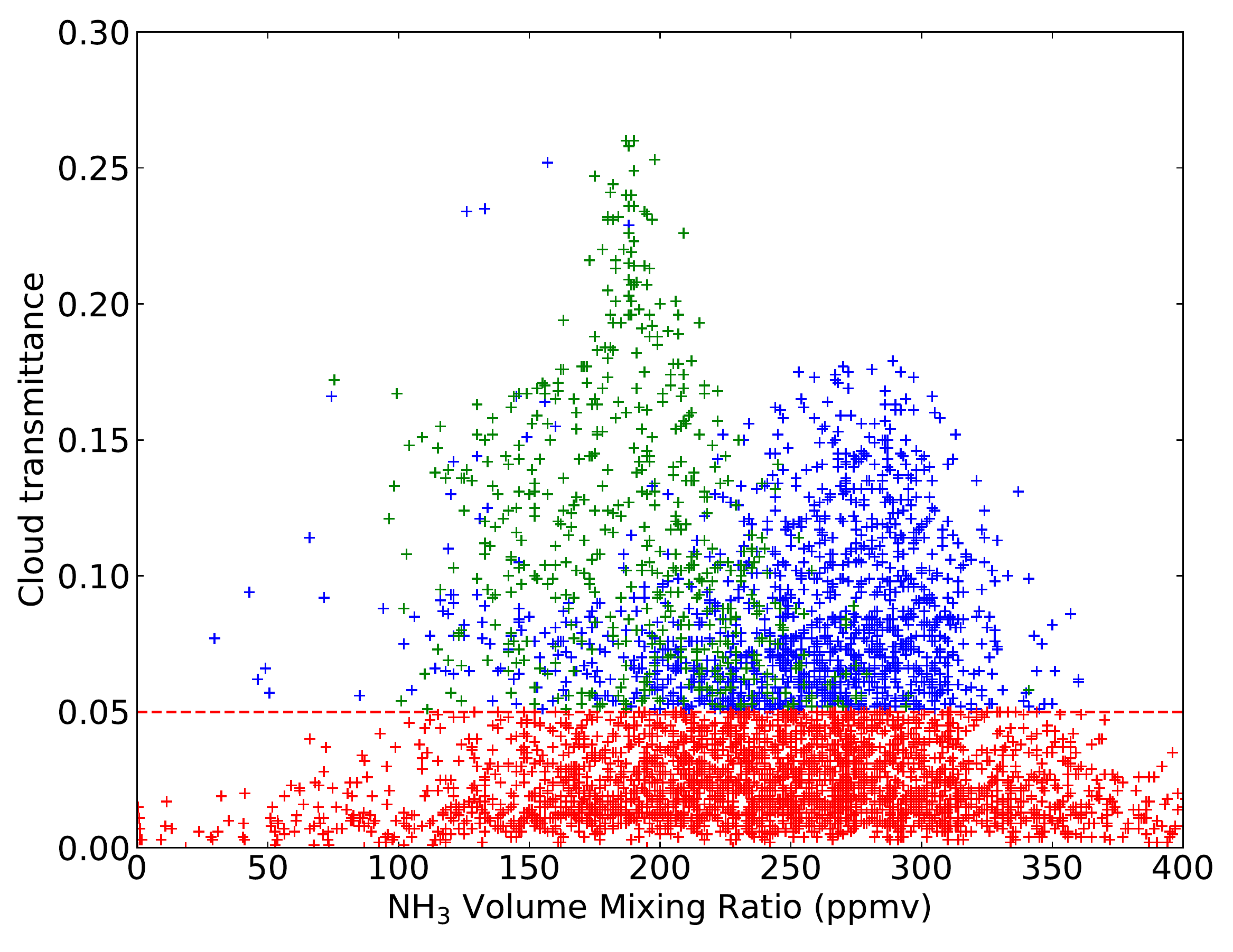}}
\subfloat[\label{subfig:nh3_vmr-cloud_transmittance_111000}]{\includegraphics[width=0.32\textwidth]{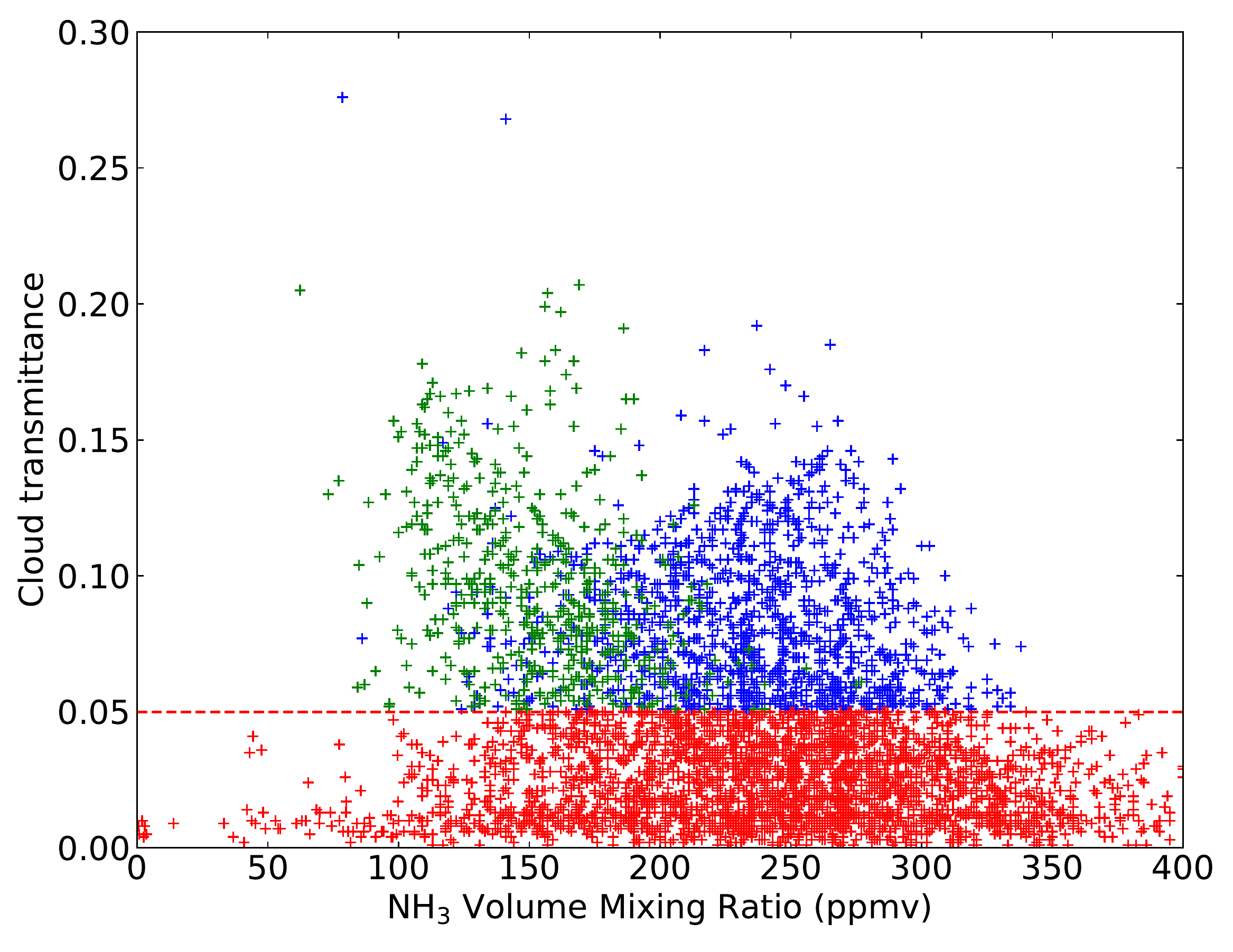}}
\subfloat[\label{subfig:nh3_vmr-cloud_transmittance_125000}]{\includegraphics[width=0.32\textwidth]{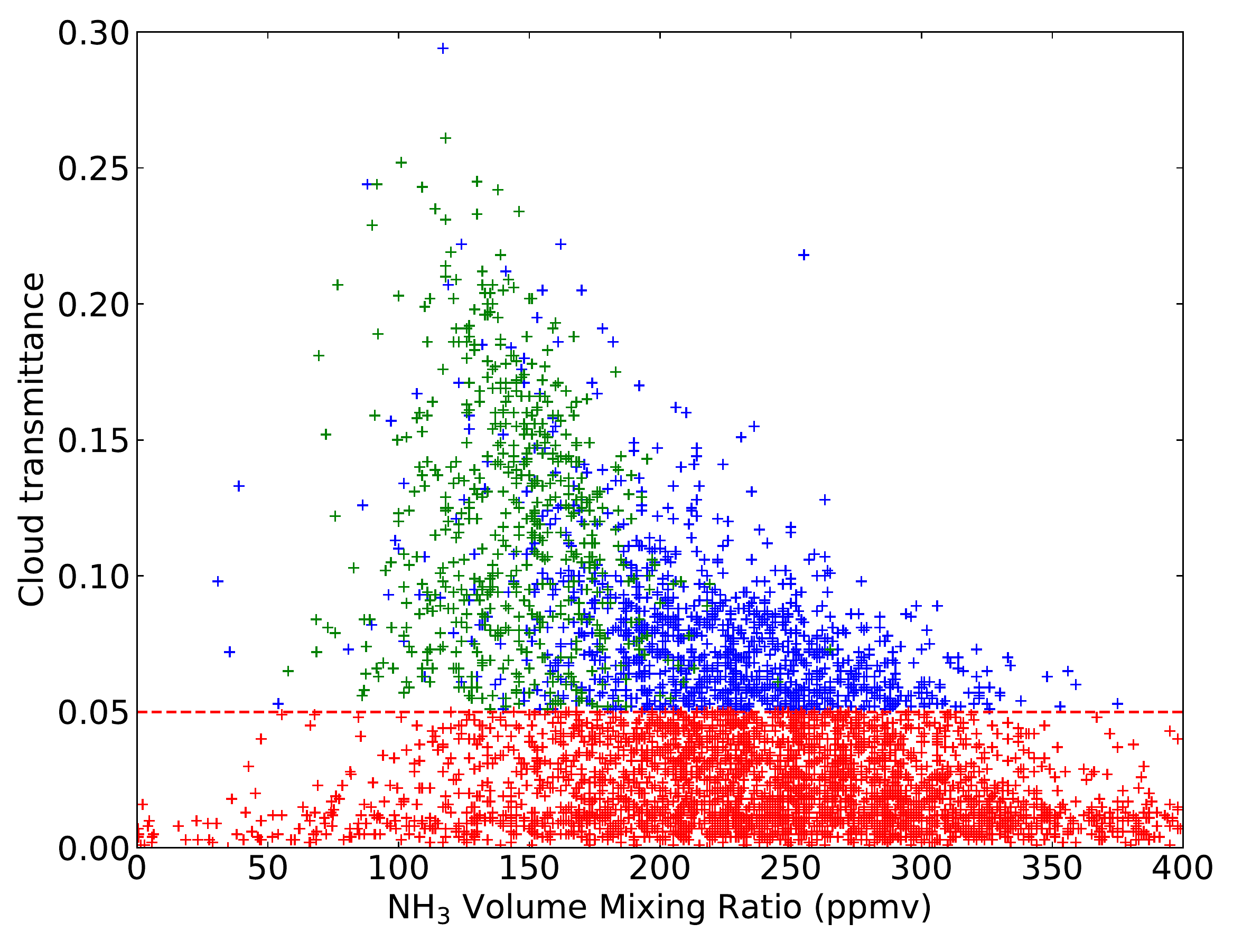}}
\\
\subfloat[\label{subfig:nh3_vmr-cloud_transmittance_nh3_vmr_2bar_093600_zones}]{\includegraphics[width=0.32\textwidth]{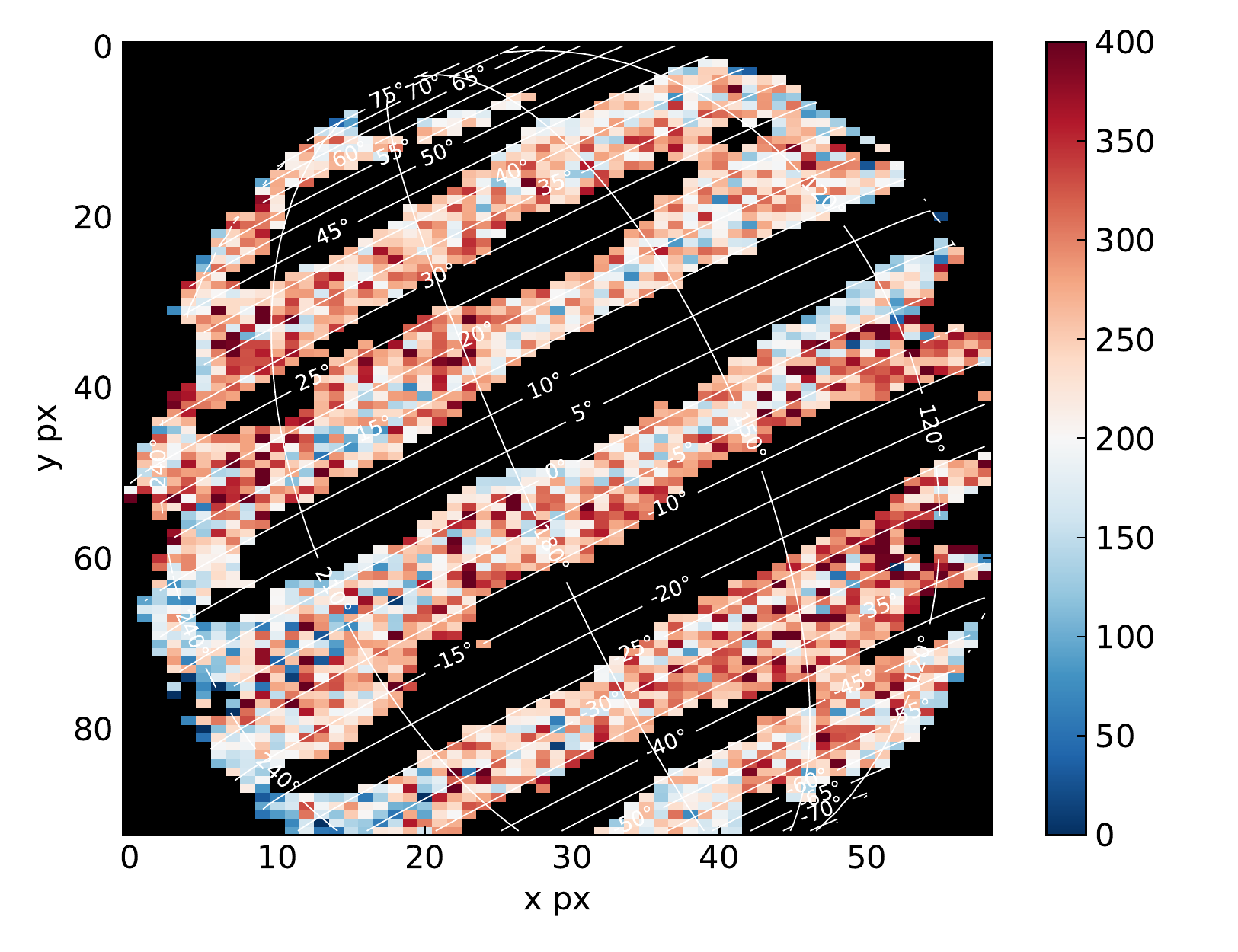}}
\subfloat[\label{subfig:nh3_vmr-cloud_transmittance_nh3_vmr_2bar_111000_zones}]{\includegraphics[width=0.32\textwidth]{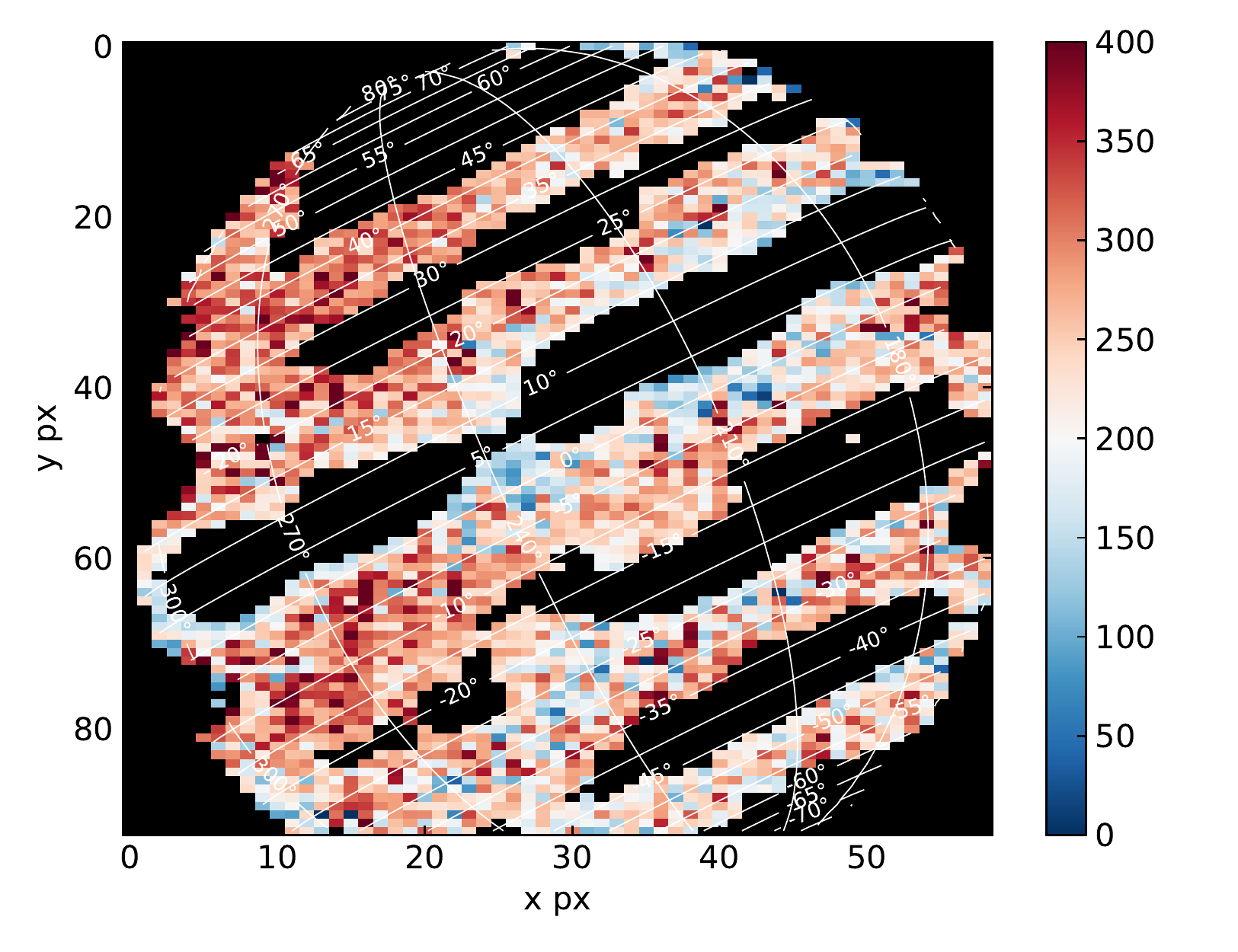}}
\subfloat[\label{subfig:nh3_vmr-cloud_transmittance_nh3_vmr_2bar_125000_zones}]{\includegraphics[width=0.32\textwidth]{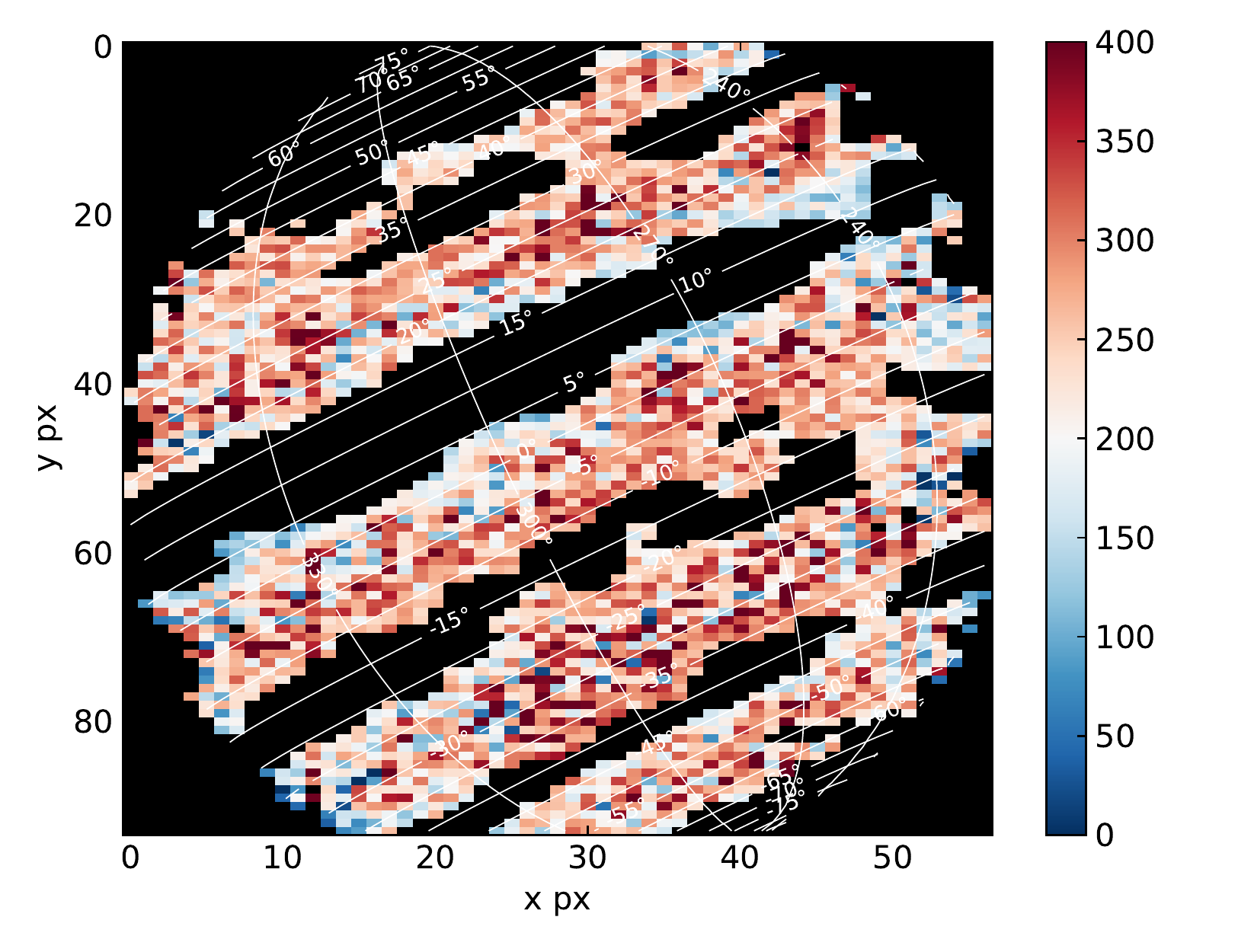}}
\\
\subfloat[\label{subfig:nh3_vmr-cloud_transmittance_nh3_vmr_2bar_093600_neb}]{\includegraphics[width=0.32\textwidth]{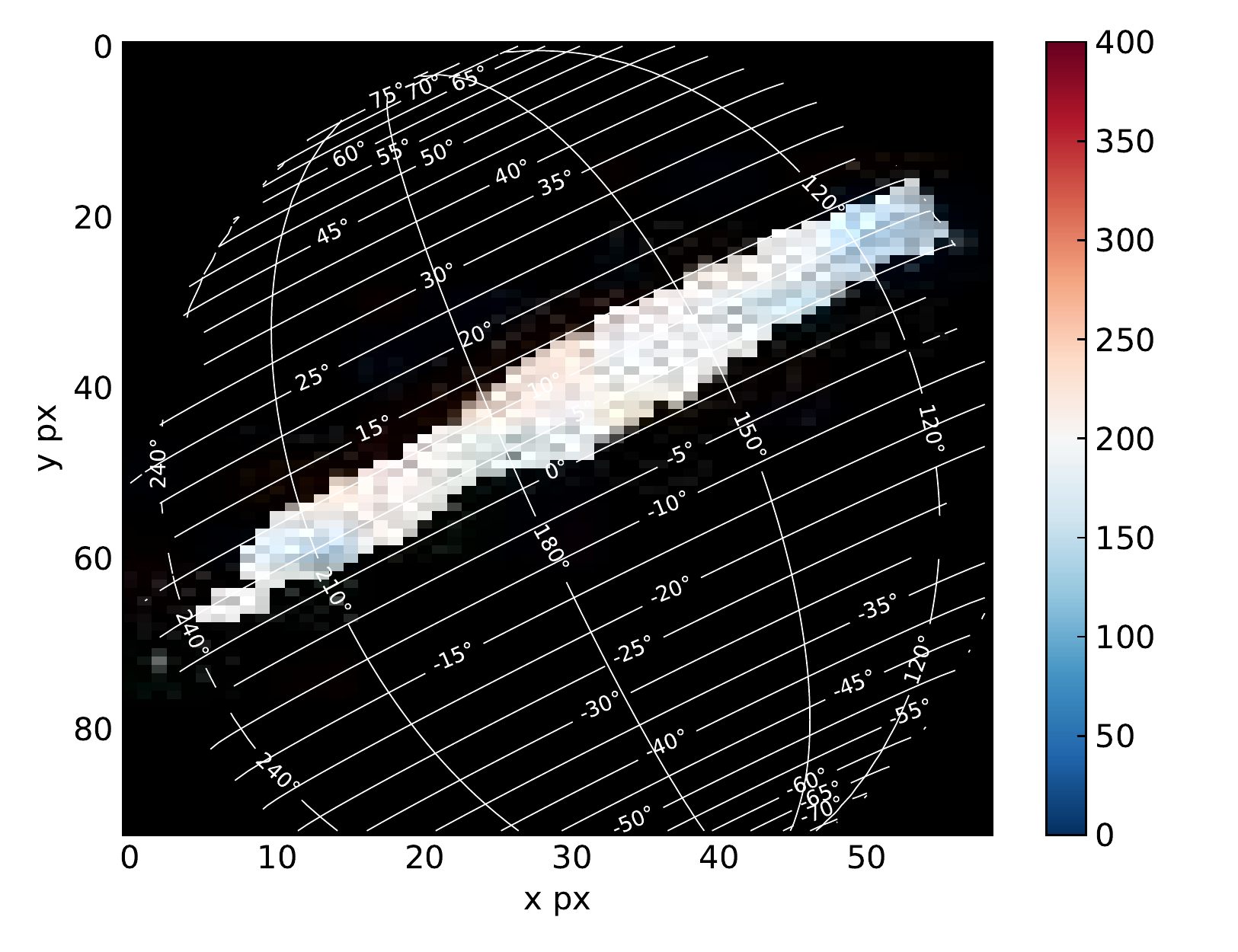}}
\subfloat[\label{subfig:nh3_vmr-cloud_transmittance_nh3_vmr_2bar_111000_neb}]{\includegraphics[width=0.32\textwidth]{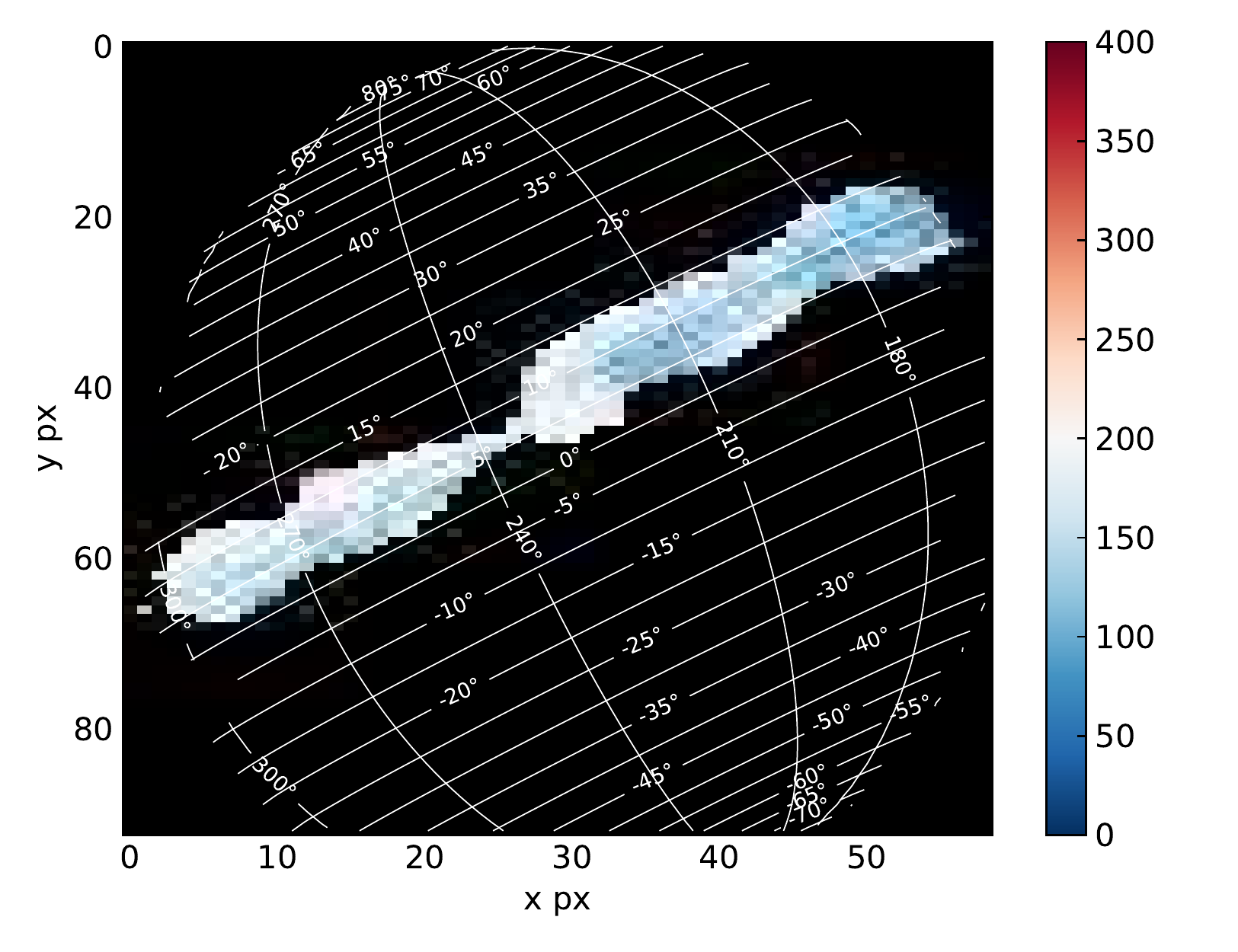}}
\subfloat[\label{subfig:nh3_vmr-cloud_transmittance_nh3_vmr_2bar_125000_neb}]{\includegraphics[width=0.32\textwidth]{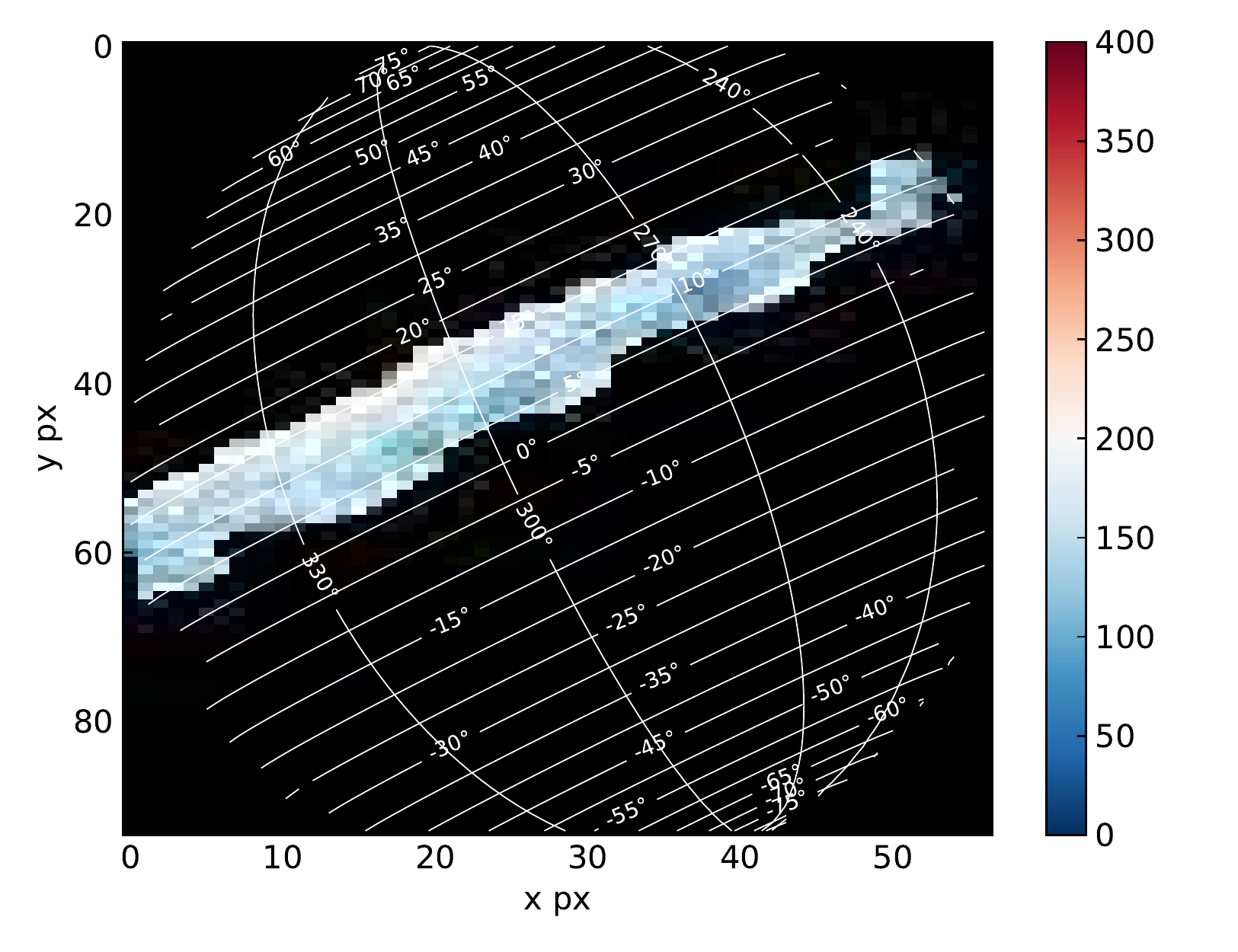}}
\\
\subfloat[\label{subfig:nh3_vmr-cloud_transmittance_nh3_vmr_2bar_093600_belts}]{\includegraphics[width=0.32\textwidth]{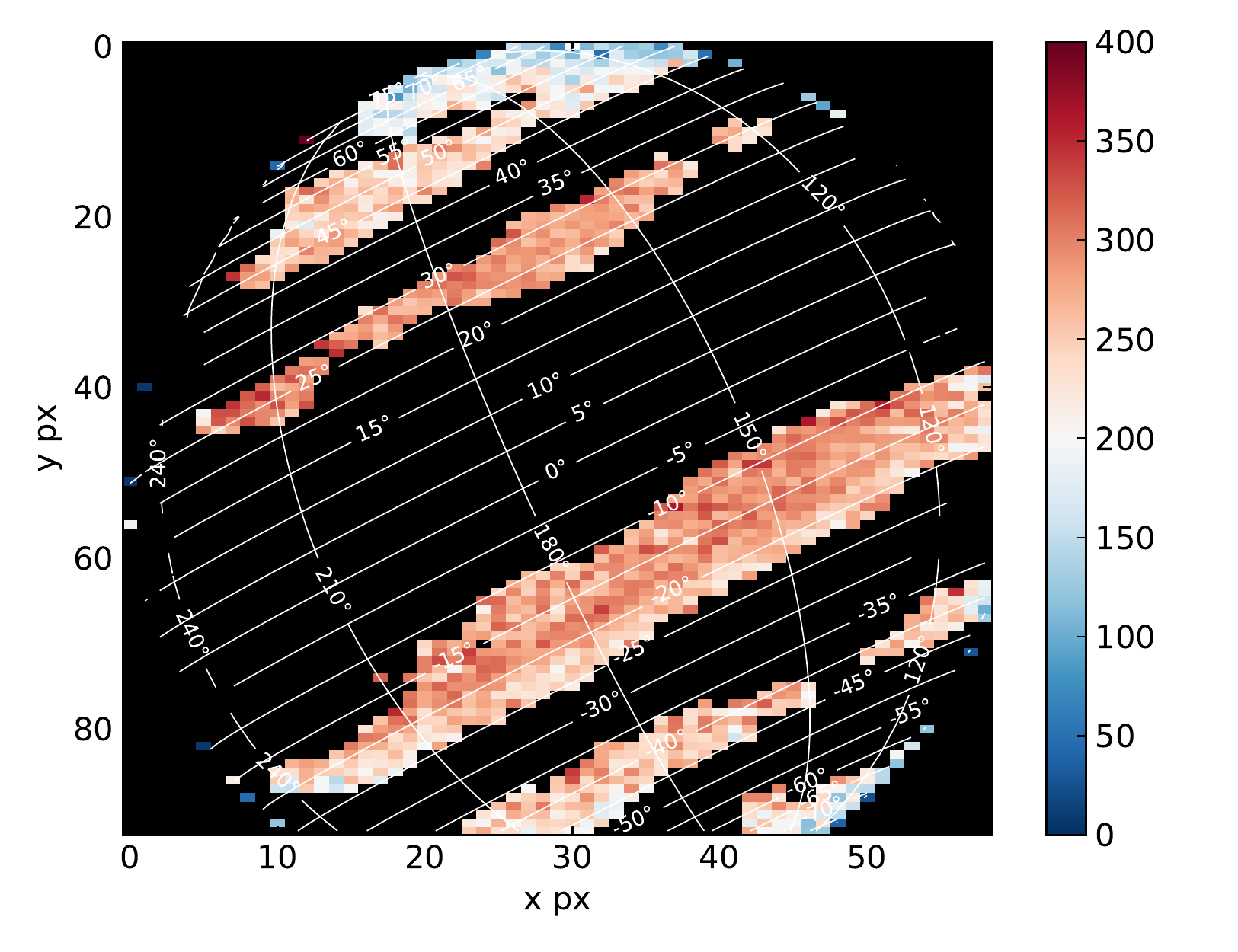}}
\subfloat[\label{subfig:nh3_vmr-cloud_transmittance_nh3_vmr_2bar_111000_belts}]{\includegraphics[width=0.32\textwidth]{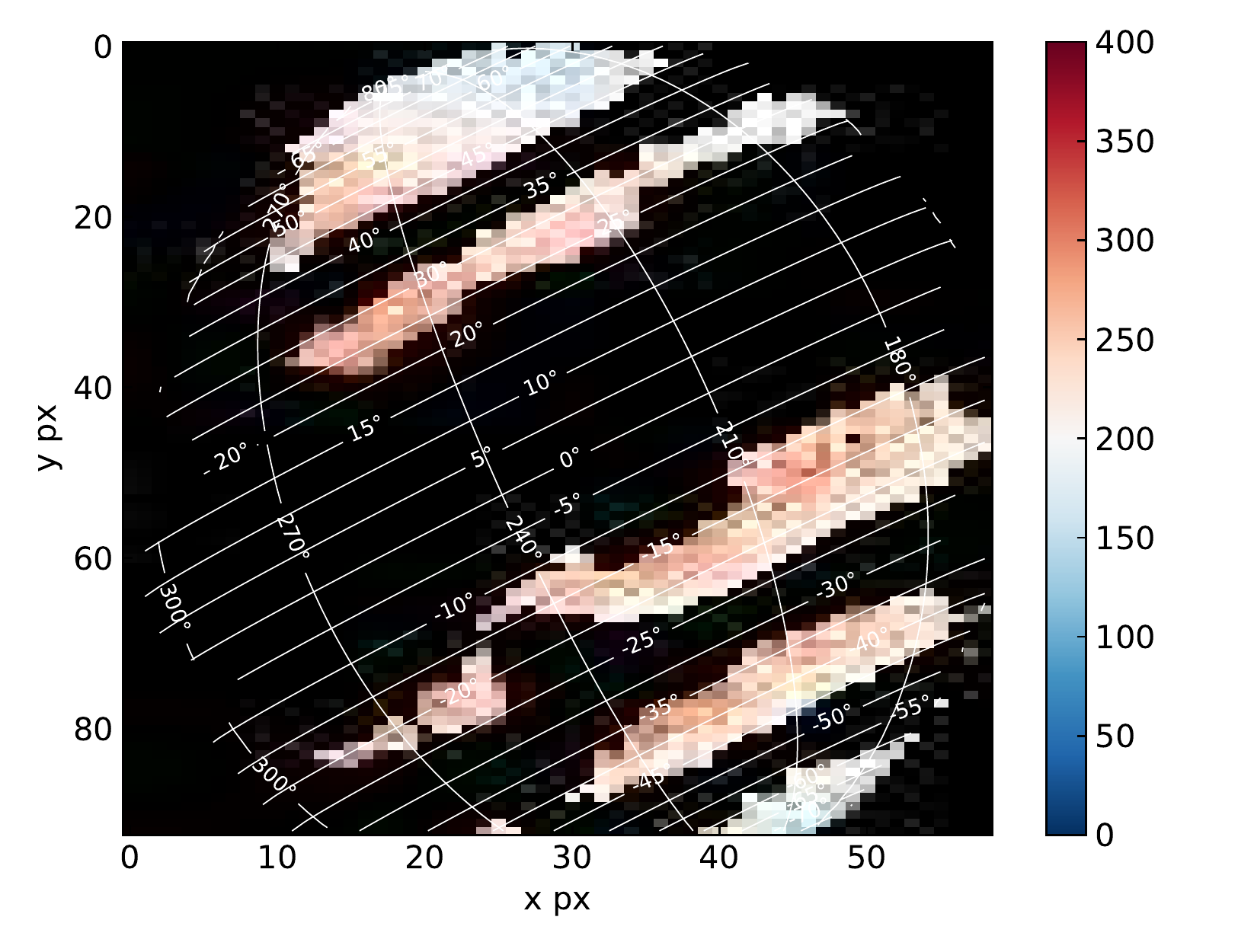}}
\subfloat[\label{subfig:nh3_vmr-cloud_transmittance_nh3_vmr_2bar_125000_belts}]{\includegraphics[width=0.32\textwidth]{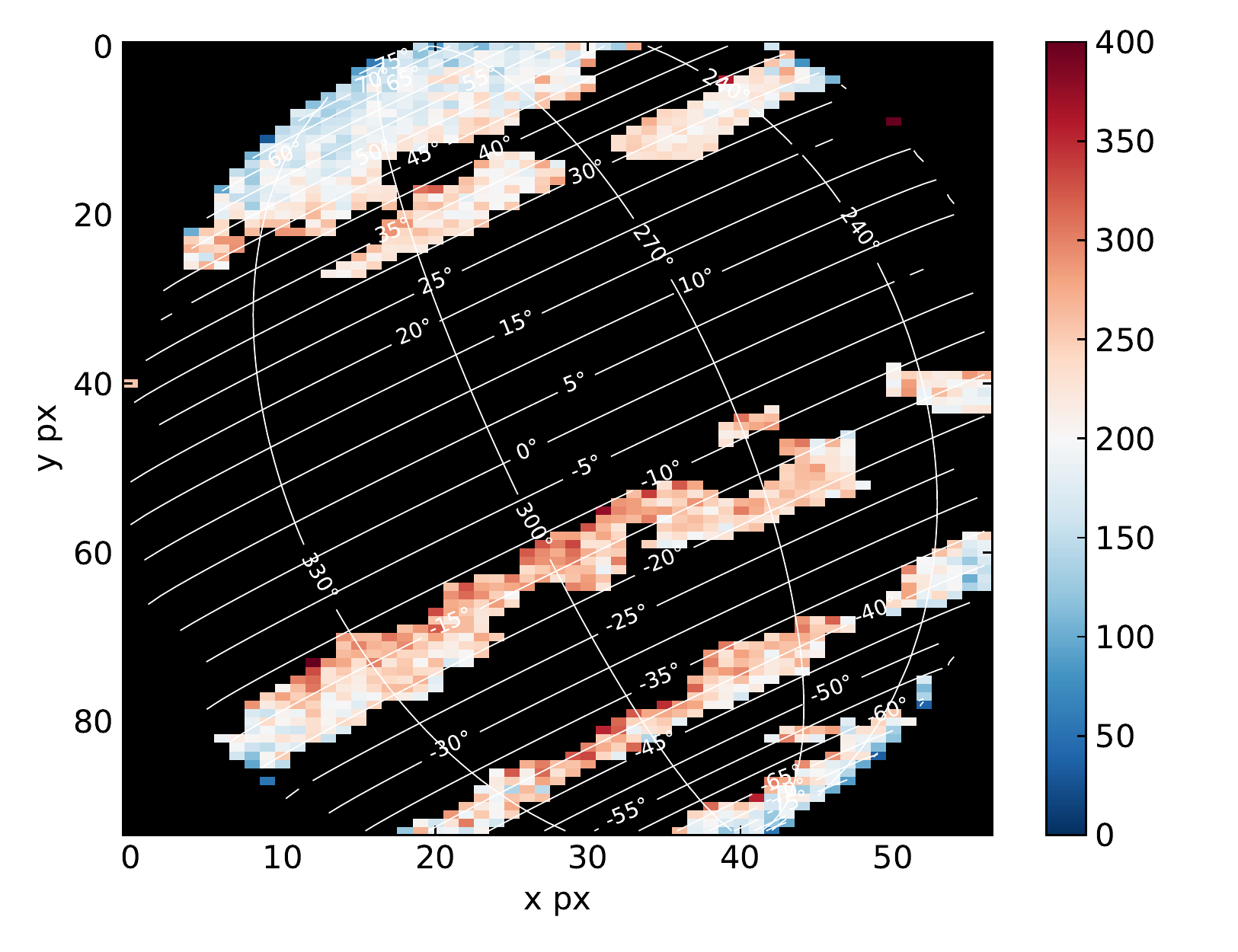}}

\caption{\label{fig:nh3_vmr-cloud_transmittance} Figures (a) to (c): retrieved cloud transmittance in function of our retrieved NH$_3$ abundance for each spectrum of the spectral cube taken at 09h36, 11h10 and 12h50 (see Table~\ref{tab:spectral_cube_parameters}) in this order from left to right. Red: spectra with a retrieved cloud transmittance lower than 0.05, corresponding to the zones. Green: spectra with a retrieved cloud transmittance greater than 0.05 and located between latitudes 0 and 17$^\circ$N, corresponding to the NEB. Blue: spectra located outside the NEB with a cloud transmittance greater than 0.05. Figures (d) to (l): retrieved NH$_3$ abundance and location of each spectra of each of our spectral cube, for each of the filters used in figure (a) to (c). From top to bottom: spectra corresponding to respectively the red, green and blue points of figures (a) to (c). From left to right: spectra corresponding to respectively the figure (a), (b) and (c). Latitudes are planetocentric and longitudes are in system III.}
\end{figure*}

\begin{figure}[t]
\centering
\includegraphics[width=0.49\textwidth]{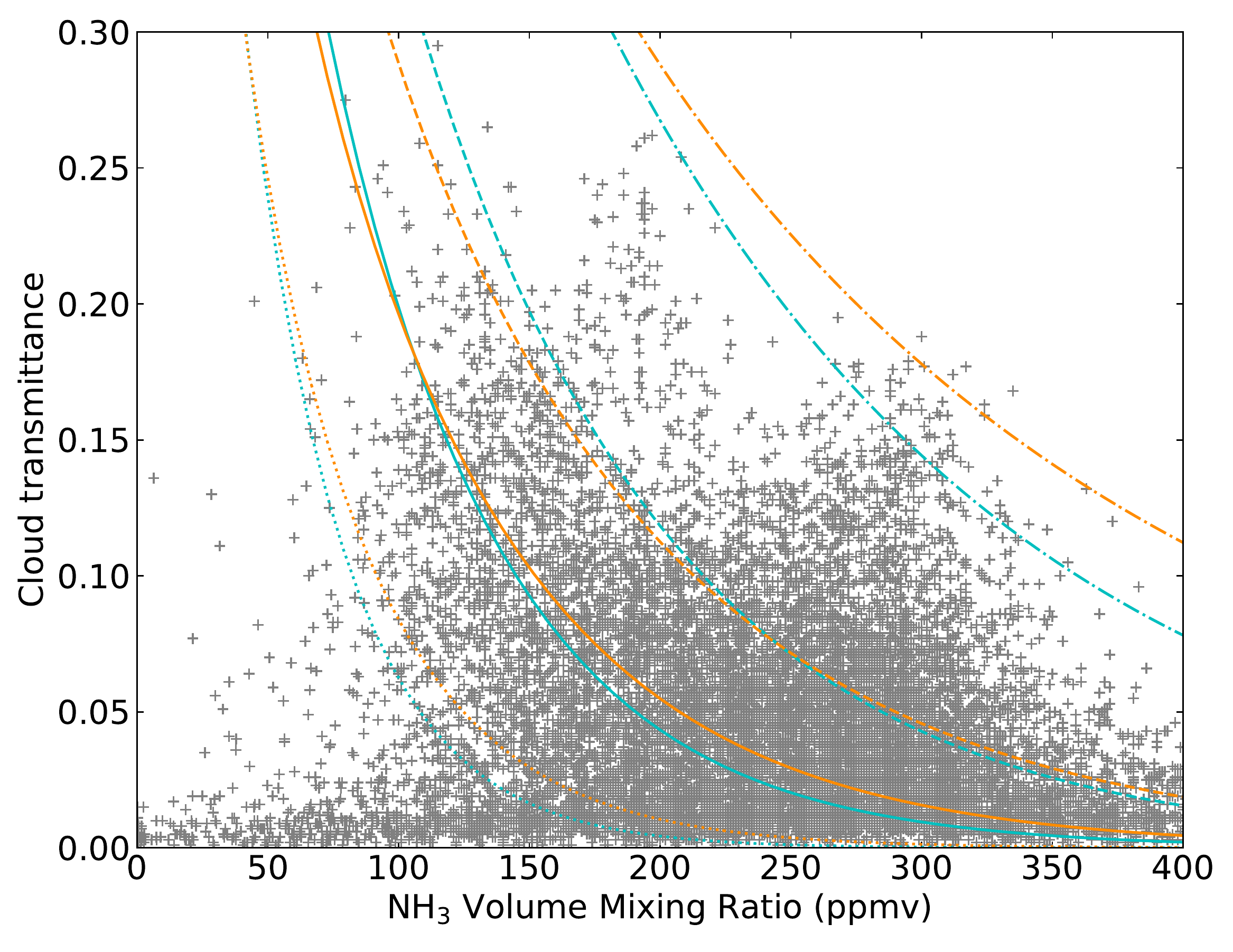}
\caption{\label{fig:nh3_vmr-cloud_transmittance_lacis1974_example} Relation between cloud transmittance and NH$_3$ VMR. Grey: retrieved cloud transmittance of each spectrum of our spectral cubes with respect to the retrieved NH$_3$ abundance at 2 bar. Blue: each line corresponds to transmittances calculated through \citet{Lacis1974}'s and \citet{Ackerman2001} cloud model for a NH$_3$-ice cloud situated at 0.8 bar and constituted of particles with a size of 10 $\mu$m, for a $f_{\text{rain}}$ of 45 (dotted), 80 (solid), 120 (dashed) and 200 (dotted-dashed). Orange: same as blue, but for a NH$_4$SH cloud situated at 1.2 bar, for a $f_{\text{rain}}$ of 150 (dotted), 250 (solid), 350 (dashed) and 700 (dotted-dashed).}
\end{figure}

In figure~\ref{fig:nh3_vmr-cloud_transmittance} we display the retrieved cloud transmittance as a function of our retrieved NH$_3$ abundance at 2 bar for each of our spectra and each of our spectral cubes. We decided to classify the spectra into three families, represented on the figures by different colors. (i) The first family (red) corresponds to the zones (clouds transmittance $<$ 0.05). (ii) The second family (green) corresponds to all the spectra outside of the zones with a latitude comprised between 0 and 17 $^\circ$N, which correspond to the NEB. (iii) The third family (blue) are all the spectra outside of the zones and NEB, which corresponds to the other belts. The location of these families are displayed in this order from top to bottom in the column below their corresponding figure. As expected, the retrieved NH$_3$ abundance at 2 bar is very dispersed in the zones, due to the low SNR. It appears that the NEB has a behaviour very different from the other belts, with generally a lower NH$_3$ abundance and a higher cloud transmittance. The asymmetry of the SEB is also visible: as we pass over the GRS from East to West (figures \ref{subfig:nh3_vmr-cloud_transmittance_093600} to \ref{subfig:nh3_vmr-cloud_transmittance_125000}), thicker clouds cover the SEB while the NH$_3$ abundance remains constant.

In figure~\ref{fig:nh3_vmr-cloud_transmittance_lacis1974_example},
we display the cloud transmittance as a function of the NH$_3$ abundance for each of our spectra. To estimate the Spearman correlation coefficient of the two parameters (i.e. how well the relationship between the two parameters can be described using a monotonic function) and the reliability of this coefficient, we applied the following methodology. (i) We generated 1000 sets of Gaussian random NH$_3$ abundances, using as mean value our retrieved NH$_3$ abundances, and as standard deviation our NH$_3$ abundance uncertainties. (ii) Then, we calculated the mean and the standard deviation of the Spearman correlation coefficient between each of those set and our retrieved cloud transmittance. Taking only the points outside of the zones (5246 points), we obtained a Spearman correlation coefficient of $\approx -$0.23 $\pm$ 0.01. We interpret these numbers as a fairly reliable evidence (low standard deviation) of a weak anti-correlation (negative near zero coefficient) between the cloud transmittance and the NH$_3$ abundance. To physically explain this correlation, we considered the cloud optical thickness model by \citet{Ackerman2001} and the multi-scattering cloud transmittance model from \citet{Lacis1974} we obtain the following relation between NH$_3$ abundance and the cloud transmittance:
\begin{eqnarray}
\tau &=& \frac{3}{2} \frac{\epsilon_p p q_{\text{NH}_3}}{g r_{\text{eff}} \rho_p (1 + f_{\text{rain}})}\\
t_c(q_{\text{NH}_3}) &=& \frac{4u}{(u + 1)^2 e^t - (u - 1)^2 e^{-t}}
\end{eqnarray}
With:
\begin{eqnarray}
u &=& \sqrt{ \frac{1 - g_c \tilde{\omega}}{1 - \tilde{\omega}} } \nonumber \\
t &=& \tau\sqrt{ 3 (1 - g_c\tilde{\omega})(1 - \tilde{\omega}) } \nonumber
\end{eqnarray}
And with $\tau$ the optical thickness of the cloud, $\epsilon_p$ the ratio of the molar mass of the particle over the molar mass of the atmosphere, $p$ the pressure of the cloud base, $q_{\text{NH}_3}$ the abundance of NH$_3$, $g$ the gravity of the planet, $r_{\text{eff}}$ the area-weighted mean particle effective radius --- which follows a log-normal distribution with a geometric standard deviation of 1.6 ---, $\rho_p$ the density of the particle, $f_{\text{rain}}$ the ratio of the mass-averaged particle sedimentation velocity to convective velocity, $g_c$ the asymmetry factor of the cloud particles, and $\tilde{\omega}$ their single-scattering albedo. The parameters $g_c$ and $\tilde{\omega}$ are calculated through Mie theory from optical constants from \citet{Howett2007} and \citet{Martonchik1984}, and are displayed in Table~\ref{tab:cloud_parameters}.

With this relation, we are able to fit the observed correlation with both an NH$_3$-ice and an NH$_4$SH cloud, for different sets of our free parameters $r_{\text{eff}}$ and $f_{\text{rain}}$ (see Table~\ref{tab:cloud_parameters}). These sets of parameters are roughly consistent with the values retrieved by \citet{Ohno2017} for example.

\begin{table*}
\centering
	\caption{\label{tab:cloud_parameters}Cloud parameters fitting the observed cloud transmittance-NH$_3$ VMR at 2 bar correlation.}
		\begin{threeparttable}
		\begin{tabular}{c c c c c c c}
			\hline
			Particle & Cloud base & Particle  & \multirow{2}{*}{$g_c$\tnote{(1)}} & \multirow{2}{*}{$\tilde{\omega}$\tnote{(1)}} & $f_{\text{rain}}$\tnote{(2)} & $K_{zz}$ (min--max)\tnote{(2)}\\
			type & pressure (bar) &  size ($\mu$m) & & & (min--max) & (cm$^2$.s$^{-1}$)\\
			\hline
			\multirow{2}{*}{NH$_3$-ice} & \multirow{2}{*}{0.8} & 10 & 0.775 & 0.923 & 45--200 &  9.1$\times$10$^5$--2.0$\times$10$^5$\\
			 & & 100 & 0.913 & 0.697 & 10--50 & 4.1$\times$10$^8$--0.8$\times$10$^8$\\
			 \multirow{2}{*}{NH$_4$SH} & \multirow{2}{*}{1.2} & 10 & 0.567 & 0.968 & 150--700 & 4.4$\times$10$^5$--0.9$\times$10$^5$\\
			 & & 100 & 0.697 & 0.867 & 25--100 & 2.6$\times$10$^8$--0.7$\times$10$^8$\\
			\hline
		\end{tabular}
		\begin{tablenotes}
		\item[(1)] At 5 $\mu$m.
		\item[(2)] The minimum and maximum values correspond to a fit close to respectively the dotted curve and the dashed-dotted curve of figure~\ref{fig:nh3_vmr-cloud_transmittance_lacis1974_example}.
		\end{tablenotes}
		\end{threeparttable}
\end{table*}

If we take large enough particles, we can use the droplet terminal fallspeed equation from \citet{Ackerman2001} to calculate the corresponding eddy diffusion coefficient $K_{zz}$. We obtain values between $\approx10^5$ and $\approx10^8$ cm$^2\cdot$s$^{-1}$, depending on the particle size we choose, for either NH$_3$ or NH$_4$SH particles. The values for 100 $\mu$m particles are of the same order of magnitude as the eddy mixing coefficient expected from free convection, as discussed by \cite{Bezard2002}. We can also note that \citet{Flasar1978}, using a model of turbulent convection in a rotating body, estimated $K_{zz}$ $\approx$ 4$\times$10$^8$ cm$^2\cdot$s$^{-1}$ at 10 bar decreasing to 1$\times$10$^8$ cm$^2\cdot$s$^{-1}$ at 0.5 bar in the NEB, in good agreement with our finding.

However, the cloud parameters we retrieved may vary for each region. For example, it is possible that the hotspots may have thinner clouds of smaller particles similar to Galileo's probe results \citep[][particle sizes $<$10 $\mu$m]{Ragent1998}, while the zones may be covered with thicker clouds of larger particles. Moreover, we have no information about the number of different cloud layers and their respective base pressure or the chemical composition of the clouds. More importantly, even if our simple model fits relatively well the behaviour of the NEB, it cannot explain the behaviour of the other belts, mainly the SEB. This inadequacy could be explained by the simplicity of our model and the complexity of the meteorology of the belts, or simply by the fact that the NH$_3$ abundance is roughly constant in the belts and is not correlated with the cloud transmittance. In particular, the use of a single eddy mixing coefficient to parametrize vertical transport is likely too simplistic as it does not account for vertical advection (either upwelling or downwelling) nor for horizontal advection.

If we assume that NH$_3$ drives at least partially the clouds on Jupiter, so if we assume that our model is adequate but too simple, our simplistic model suggests that both NH$_3$-ice and NH$_4$SH are cloud constituents compatible with our observations, but also that local meteorology probably plays a significant in distributing gas and cloud particles.

\section{Conclusions}
We were able to derive a map of the abundance of NH$_3$ on Jupiter at 2 bar with a spatial resolution of $\approx$ 0.7'' and a mean uncertainty of 20$\%$. The latitude coverage is 75$^\circ$S--75$^\circ$N (planetocentric) and the longitude coverage is 90--360$^\circ$W (system III), although the information we retrieved in the zones is of questionable quality.

Our results notably show:
\begin{enumerate}
\item a large NH$_3$ depletion compared to other regions in the NEB, with mixing ratio values around 160 ppmv at 2 bar, of varying width with longitude,
\item a correlation between the most depleted regions in NH$_3$ of the disk, the brightest regions at 5 $\mu$m, the "hotspots", and the blue-gray regions in the NEB in visible light,
\item an enrichment in NH$_3$ ($\approx$ 250 $\pm$ 50 ppmv at 2 bar) in the other belts compared to the NEB, with some local exceptions (notably in the NTB and around the GRS),
\item a north-to-south NH$_3$ abundance "slope" in the belts, at the exception of the NEB, with a poleward orientation,
\item a similar north-to-south NH$_3$ abundance slope northward of latitude $\approx$ 45$^\circ$N, and possibly southward of latitude $\approx$ 50$^\circ$S, starting at $\approx$ 240 $\pm$ 50 ppmv and going down to $\approx$ 160 $\pm$ 30 ppmv,
\item in the zones and particularly in the EZ, our data suggest a NH$_3$ abundance increasing with depth, from 100 $\pm$ 15 ppmv at 1 bar to 500 $\pm$ 30 ppmv at 3 bar, albeit with a relatively low confidence level,
\item a clear distinction in the behaviour of the NEB and the other belts, in term of NH$_3$ abundance and cloud transmittance,
\item a strong dichotomy in term of cloud transmittance in the SEB, between the East and the West of the GRS: the western side seems to be perturbed by the GRS and is covered by thicker clouds than the eastern side,
\item a possible correlation between cloud transmittance and NH$_3$ abundance, more obvious in the NEB than in the other belts,
\item according to cloud models, the above mentioned correlation seems to indicate that NH$_3$ plays a major role in the formation of the cloud we observe at 5 $\mu$m, either through NH$_3$-ice or NH$_4$SH particles. However our observations also shows that local meteorology probably play an important role in cloud formation.
\end{enumerate}

Compared with MWR results \citep{Bolton2017}, we obtain on average a similar NH$_3$ abundance in the 1--3 bar region ($\approx$ 250 ppmv), and we observe the same depletion in the NEB (down to $\approx$ 150 $\pm$ 30 ppmv at 2 bar). The behaviour of NH$_3$ at pressures greater than 4 bars is inaccessible to us, due to the lack of sensitivity of our spectral range to greater pressures. We do not observe the "plume" in the EZ and the behaviour of the zones in general seems to be different, but the results we obtain in this region are not as reliable as those for the belts.

\section*{Acknowledgements} 
D.B. and T.F. were supported by the Programme National de Planétologie as well as the Centre National d'Etudes Spatiales. 

T.G. acknowledges funding supporting this work from NASA PAST through grant number NNH12ZDAO01N-PAST.

G.S.O. was supported by funds from the National Aeronautics and Space Administration distributed to the Jet Propulsion Laboratory, California Institute of Technology.

Fletcher was supported by a Royal Society Research Fellowship at the University of Leicester.

\section*{References}

\bibliographystyle{myplainnat}

\bibliography{texes_article_2018}

\onecolumn
\section*{Annexes}
\label{sec:annexes}
These annexes describe the equations used in our radiative transfer model.

\subsection{Optical depth}
Relationship between $c_m(z)$, the column number density of attenuating specie $m$ at altitude $z$ and $\mathcal{V}_m(z)$, the Volume Mixing Ratio (a.k.a. abundance) of specie $m$ at altitude $z$:
\begin{equation}
c_m(z) = \int_0^\infty \mathcal{V}_m(z) n_0[T(z), P(z)] dz
\end{equation}
where $n_0[T(z), P(z)]$ is the Loschmidt constant at temperature $T(z)$ and pressure $P(z)$.

Optical depth at wavenumber $\nu$, atmospheric level $k$ and for attenuating specie $m$:
\begin{equation}
\tau_{\nu, m}(k) = \sum_{l=k}^{N_{\text{levels}}} \sigma_m(\nu, l) c_m(l) = \sum_{l=k}^{N_{\text{levels}}} \alpha_m(\nu, l)
\end{equation}
where $N_{\text{levels}}$ is the number of atmospheric levels and $\sigma_m$ is the attenuation cross section of the attenuating specie $m$. $\alpha_m(\nu, l)$ will be called "column attenuation coefficient", with $\alpha_m(\nu, l) \equiv \sigma_m(\nu, l) c_m(l)$.

Optical depth at wavenumber $\nu$ and for atmospheric level $k$:
\begin{equation}
\tau_{\nu}(k) = \sum_{m=1}^{N_{\text{species}}} \tau_{\nu, m}(k)
\end{equation}
where $N_{\text{species}}$ is the number of different attenuating species.

Optical depth at cloud level $k_c$:
\begin{equation}
\tau_{\nu}(k_c) = \sum_{m=1}^{N_{\text{species}}} \sum_{l=k_c}^{N_{\text{levels}}} \alpha_m(\nu, l)
\end{equation}

Relationship between optical depth and spectral directional transmittance at emission angle $\theta_e$:
\begin{equation}
\mathcal{T}_{\nu}(k, \theta_e) = e^{-\tau_{\nu}(k)\sec(\theta_e)}
\end{equation}

Derivative of spectral directional transmittance over logarithm of column number density of specie $m$:
\begin{equation}
\begin{alignedat}{2}
&&\frac{\partial \mathcal{T}_{\nu}(k, \theta_e)}{\partial \ln c_m(k)} &= \frac{\partial \exp\left(-\sec(\theta_e) \sum_{s=1}^{N_{\text{species}}} \sum_{l=k}^{N_{\text{levels}}} \sigma_{s}(\nu, l) e^{\ln c_s(l)} \right)}{\partial \ln c_m(k)} \\
&\Leftrightarrow &\frac{\partial\mathcal{T}_{\nu}(k, \theta_e)}{\partial \ln c_m(k)} &= -\sec(\theta_e) \alpha_m(\nu, k) e^{-\tau_{\nu}(k)\sec(\theta_e)}
\end{alignedat}
\end{equation}

\subsection{Cloud transmittance}
Total cloud hemispherical transmittance:
\begin{equation}
\mathcal{T}_{c}(k) = \prod_{l=k}^{N_{\text{levels}}}\left[\prod_{n=1}^{N_{\text{clouds}}} \mathcal{T}_{c, n}(k)\right]
\end{equation}
where $\mathcal{T}_{c, n}(k)$ is the hemispherical transmittance of cloud $n$ at level $k$. $\mathcal{T}_{c, n}(k) = 1$ above the cloud top level and below the cloud bottom level.

Derivative of total cloud hemispherical transmittance over cloud $n$ hemispherical transmittance:
\begin{equation}
\begin{alignedat}{2}
&&\frac{\partial \mathcal{T}_c(k)}{\partial \mathcal{T}_{c, n}(k)} &= \frac{\partial \prod_{l=k}^{N_{\text{levels}}}\left[\prod_{n=1}^{N_{\text{clouds}}} \mathcal{T}_{c, n}(l)\right]}{\partial \mathcal{T}_{c, n}(k)} \\
&\Leftrightarrow &\frac{\partial \mathcal{T}_c(k)}{\partial \mathcal{T}_{c, n}(k)} &= \frac{\mathcal{T}_c(k)}{\mathcal{T}_{c, n}(k)} \\
&\Leftrightarrow &\frac{\partial \sum_{l=1}^{N_{\text{levels}}} \mathcal{T}_{c}(l)}{\partial \mathcal{T}_{c, n}(k)} &= \frac{\mathcal{T}_{c}(1)}{\mathcal{T}_{c, n}(k)} + \frac{\mathcal{T}_{c}(2)}{\mathcal{T}_{c, n}(k)} + ... + \frac{\mathcal{T}_{c}(k)}{\mathcal{T}_{c, n}(k)} + 0 + ... + 0 \\
&\Leftrightarrow &\frac{\partial \sum_{l=1}^{N_{\text{levels}}} \mathcal{T}_{c}(l)}{\partial \mathcal{T}_{c, n}(k)} &= \frac{\sum_{l=1}^{k} \mathcal{T}_{c}(l)}{\mathcal{T}_{c, n}(k)}
\end{alignedat}
\end{equation}

\subsection{Reflection contribution}
Spectral radiance reflected by the cloud at emission angle $\theta_e$:
\begin{equation}
\begin{aligned}
L_r(\nu) = \mathcal{R}_c E_\odot(\nu) e^{-\sec(\theta_\odot)\tau_{\nu}(k_c)} e^{-\sec(\theta_e)\tau_{\nu, m}(k_c)}
\end{aligned}
\end{equation}
with $\mathcal{R}_c$ the hemispherical reflectance of the cloud, $E_\odot$ the spectral irradiance of the Sun and $\theta_\odot$ the cosine of the local zenith angle. Level $k_c$ is the atmospheric level of the highest altitude cloud.

Derivative of radiance reflected by the cloud over logarithm of gas abundance:
\begin{equation}
\begin{aligned}
&&\frac{\partial L_r}{\partial \ln c_m(k)} &= \frac{\partial \mathcal{R}_c E_\odot e^{-\left[\sec(\theta_e) + \sec(\theta_\odot)\right]\tau_{\nu}(k_c)}}{\partial \ln c_m(k)} && \\
&\Leftrightarrow &\frac{\partial L_r}{\partial \ln c_m(k)} &= -\left[\sec(\theta_e) + \sec(\theta_\odot)\right] \sigma_m(\nu, k) c_m(k) L_r & \text{if $k \geq k_c$, 0 else.} &
\end{aligned}
\end{equation}

Derivative of spectral radiance reflected by the cloud over cloud $n$ hemispherical transmittance at level $k$:
\begin{equation}
\begin{aligned}
&&\frac{\partial L_r}{\partial \mathcal{T}_{c, n}(k)} &= \frac{\partial \mathcal{R}_c E_\odot e^{-\left[\sec(\theta_e) + \sec(\theta_\odot)\right] \tau_{\nu}(k_c)}}{\partial \mathcal{T}_{c, n}(k)} \\
&\Leftrightarrow &\frac{\partial L_r}{\partial \mathcal{T}_{c, n}(k)} &\approx \frac{\partial (1-\mathcal{T}_{c}(k_c))}{\partial \mathcal{T}_{c, n}(k)} E_\odot e^{-\left[\sec(\theta_e) + \sec(\theta_\odot)\right]\tau_{\nu}(k_c)}\\
&\Leftrightarrow &\frac{\partial L_r}{\partial \mathcal{T}_{c, n}(k)} &\approx -\frac{\mathcal{T}_{c}(k_c)}{\mathcal{T}_{c, n}(k)}E_\odot e^{-\left[\sec(\theta_e) + \sec(\theta_\odot)\right] \tau_{\nu}(k_c)} &\text{if $k \geq k_c$, 0 else.}
\end{aligned}
\end{equation}

\subsection{Thermal contribution}
Thermal spectral radiance:
\begin{equation}
\begin{aligned}
&&L_{th}(\nu) &= \int_0^\infty \mathcal{T}_c(z) B_{\nu}[T(z)] de^{-\sec(\theta_e)\tau_{\nu}(z)} \\
&\Leftrightarrow &L_{th}(\nu) &\approx \sum_{l=1}^{N_{\text{levels}}}\mathcal{T}_{c}(l)\left(e^{-\tau_{\nu}(l+ 1)\sec(\theta_e)} - e^{-\tau_{\nu}(l)\sec(\theta_e)}\right)B_{\nu}[T(l)] \\
&\Rightarrow &L_{th}(\nu) &= \mathcal{T}_{c}(1)(e^{-\tau_{\nu}(2)\sec(\theta_e)} - e^{-\tau_{\nu}(1)\sec(\theta_e)})~B_{\nu}[T(1)] + \mathcal{T}_{c}(2)(e^{-\tau_{\nu}(3)\sec(\theta_e)} - e^{-\tau_{\nu}(2)\sec(\theta_e)}) B_{\nu}[T(2)] + ... \\
&\Rightarrow &L_{th}(\nu) &= - \mathcal{T}_{c}(1) B_{\nu}[T(1)]e^{-\tau_{\nu}(1)\sec(\theta_e)} + \mathcal{T}_{c}(2)(B_{\nu}[T(1)]- B_{\nu}[T(2)])e^{-\tau_{\nu}(2)\sec(\theta_e)} + ... + \mathcal{T}_c(k)B_{\nu}[T(k)]e^{-\tau_{\nu}(k)\sec(\theta_e)}\\
&\Rightarrow &L_{th}(\nu) &= - \mathcal{T}_{c}(1)B_{\nu}[T(1)]e^{-\tau_{\nu}(1)\sec(\theta_e)} + \sum_{l=2}^{N_{\text{levels}}}\mathcal{T}_{c}(l)(B_{\nu}[T(l-1)]- B_{\nu}[T(l)])e^{-\tau_{\nu}(l)\sec(\theta_e)}
\end{aligned}
\end{equation}

Derivative of thermal spectral radiance over logarithm of column number density of specie $m$:
\begin{equation}
\begin{aligned}
&&\frac{\partial L_{th}(\nu)}{\partial \ln c_m(k)} &= \frac{\partial \left(- \mathcal{T}_{c}(1)B_{\nu}[T(1)]e^{-\tau_{\nu}(1)\sec(\theta_e)} + \sum_{l=2}^{N_{\text{levels}}} \mathcal{T}_{c}(l)(B_{\nu}[T(l-1)]- B_{\nu}[T(l)])e^{-\tau_{\nu}(l)\sec(\theta_e)} \right)}{\partial \ln c_m(k)} \\	
&\Leftrightarrow &\frac{\partial L_{th}(\nu)}{\partial \ln c_m(k)} &= \sec(\theta_e) \alpha_m(\nu, k) \left(\mathcal{T}_{c}(1) B_{\nu}[T(1)] e^{-\tau_{\nu}(1)\sec(\theta_e)} + \sum_{l=2}^{k} \mathcal{T}_{c}(l)(B_{\nu}[T(l)]- B_{\nu}[T(l-1)]) e^{-\tau_{\nu}(l)\sec(\theta_e)} \right)
\end{aligned}
\end{equation}

Derivative of thermal spectral radiance over cloud transmittance at level $k$:
\begin{equation}
\begin{aligned}
&&\frac{\partial L_{th}(\nu)}{\partial \mathcal{T}_{c, n}(k)} &= \frac{\partial \left(\sum_{l=1}^{N_{\text{levels}}}\mathcal{T}_{c}(l)(e^{-\tau_{\nu}(l+ 1)\sec(\theta_e)} - e^{-\tau_{\nu}(l)\sec(\theta_e)})~B_{\nu}[T(l)]\right)}{\partial \mathcal{T}_{c, n}(k)} \\	
&\Leftrightarrow &\frac{\partial L_{th}(\nu)}{\partial \mathcal{T}_{c, n}(k)} &= \frac{\sum_{l=1}^{k}\mathcal{T}_{c}(l)(e^{-\tau_{\nu}(l+ 1)\sec(\theta_e)} - e^{-\tau_{\nu}(l)\sec(\theta_e)})~B_{\nu}[T(k)]}{\mathcal{T}_{c, n}(k)}
\end{aligned}
\end{equation}

\subsection{Cloud contribution}
Cloud spectral radiance:
\begin{equation}
\begin{aligned}
&&L_{c}(\nu, k) &= d\mathcal{T}_c(k)/dk~e^{-\tau_{\nu}(k)\sec(\theta_e)} B_{\nu}[T(k)] \\
&\Leftrightarrow &L_{c}(\nu, k) &\approx (\mathcal{T}_c(k + 1) - \mathcal{T}_c(k)) e^{-\tau_{\nu}(k)\sec(\theta_e)} B_{\nu}[T(k)] \\
&\Leftrightarrow &L_{c}(\nu) &\approx \sum_{l=1}^{N_{\text{levels}}} (\mathcal{T}_c(l + 1) - \mathcal{T}_{c}(l))~e^{-\tau_{\nu}(k)\sec(\theta_e)} B_{\nu}[T(l)]
\end{aligned}
\end{equation}

Derivative of cloud spectral radiance over logarithm of column density number:
\begin{equation}
\begin{aligned}
&&\frac{\partial L_c(\nu)}{\partial \ln c_m(k)} &\approx \frac{\partial  \sum_{l=1}^{N_{\text{levels}}} (\mathcal{T}_c(k + 1) - \mathcal{T}_c(k)) e^{-\tau_{\tilde{\nu,k,gas}}\sec(\theta_e)} B_{\nu}[T(k)]}{\partial \ln c_m(k)} \\
&\Leftrightarrow &\frac{\partial L_c(\nu)}{\partial \ln c_m(k)} &\approx - \sec(\theta_e) \alpha_m(\nu, k) \sum_{l=1}^{k} (\mathcal{T}_c(k + 1) - \mathcal{T}_{c}(l)) e^{-\tau_{\nu}(k)\sec(\theta_e)} B_{\nu}[T(l)]
\end{aligned}
\end{equation}

Derivative of cloud spectral radiance over cloud hemispherical transmittance:
\begin{equation}
\begin{aligned}
&&\frac{\partial L_c(\nu)}{\partial \mathcal{T}_{c, n}(k)} &\approx \frac{\partial \sum_{l=1}^{N_{\text{levels}}} (\mathcal{T}_{c}(l + 1) - \mathcal{T}_{c}(l)) e^{-\tau_{\nu}(l)\sec(\theta_e)} B_{\nu}[T(l)]}{\partial \mathcal{T}_{c, n}(k)} \\
&\Leftrightarrow &\frac{\partial L_c(\nu)}{\partial \mathcal{T}_{c, n}(k)} &\approx \frac{\left[\sum_{l=1}^{k-1} (\mathcal{T}_{c}(l + 1) - \mathcal{T}_{c}(l)) e^{-\tau_{\nu}(l)\sec(\theta_e)} B_{\nu}[T(l)]\right] - \mathcal{T}_c(k) e^{-\tau_{\nu}(k)\sec(\theta_e)} B_{\nu}[T(k)]}{\mathcal{T}_{c, n}(k)}
\end{aligned}
\end{equation}

\subsection{Total contribution}
Total spectral radiance:
\begin{equation}
\begin{aligned}
L(\nu) = L_{th}(\nu) + L_{c}(\nu) + L_{r}(\nu)
\end{aligned}
\end{equation}

\end{document}